\documentclass[a4paper,12pt]{article}
\usepackage[a4paper,textwidth=420pt]{geometry}
\usepackage{jheppub}
\usepackage[T1]{fontenc}          
\usepackage[utf8]{inputenc}       
\usepackage{amsmath, amssymb, amsfonts, mathrsfs}
\usepackage{braket}
\usepackage{graphicx}
\usepackage{gensymb}
\usepackage{hyperref}
\usepackage{tikz}         
\usepackage{xcolor}       
\usetikzlibrary{shapes.misc} 
\usetikzlibrary{arrows.meta}  
\definecolor{MyGreen}{rgb}{0.1,0.7,0.1}
\usepackage{array}
\usepackage{appendix}
\usetikzlibrary{shapes.geometric}
\usetikzlibrary{positioning}
\usepackage{pgffor}
\usetikzlibrary{positioning,shapes.geometric}
\tikzset{
    fermion/.style={->,thick},
    fermionbar/.style={<-,thick},
    scalar/.style={dashed},
    scalarbar/.style={dashed}
}

\title{\Large \bf 
Maximal Abelian Flavor Symmetries 
}

\author[1,2]{Juanca Carrasco-Martinez}
\emailAdd{jc.carrasco@berkeley.edu}
\author[1,2]{and Lawrence J. Hall}
\emailAdd{ljh@berkeley.edu}
\affiliation[1]{Leinweber Institute for Theoretical Physics,
Department of Physics, University of California, Berkeley, CA 94720, USA}

\affiliation[2]{Theoretical Physics Group,
Lawrence Berkeley National Laboratory, Berkeley, California 94720, USA}

\abstract{A framework, MAFS, is introduced that provides an approximate description of the hierarchies of quark and lepton masses and mixing angles in terms of a set of small parameters, $\epsilon_a$, one for each fermion multiplet.  MAFS is an alternative to the Froggatt-Nielsen mechanism and has a unique application in any theory, as there are no fermion charges to choose. It becomes more powerful as the number of multiplets is reduced. In $SU(5)$ unified theories, 15 observed mass ratios and mixing angles are described, at the factor of two level, by five small $\epsilon_a$ parameters.  Even though quarks and leptons are unified, the observed hierarchical pattern of quark masses and mixings {\it requires} large neutrino mixing angles and small neutrino mass hierarchies. In an $SO(10)$ unified theory, MAFS successfully describes the 15 observed flavor hierarchies with just three small $\epsilon_a$, taking values of $0.01, 0.02$ and $0.002$. The observed cosmological baryon asymmetry results approximately from leptogenesis using MAFS in $SU(5)$, without the need for any additional small parameter; while in $SO(10)$, a further small parameter of about 0.2 appears necessary.}

\begin{document}

\maketitle

\newpage

\section{Introduction}

A theory of flavor should explain why there are three generations of quarks and leptons and allow a precise calculation of many of the quark and lepton masses and mixing angles. In this paper, we have a more modest aim: to provide an approximate description of the hierarchy of masses and mixing angles. 

In the Standard Model (SM), augmented by dimension-5 operators to include neutrino masses, flavor is described by three Yukawa coupling matrices and the neutrino mass matrix. In an arbitrary basis, these matrices contain 66 real parameters while describing 22 physically independent observables, 18 of which are measured. How are we to make progress on describing the structure of these matrices? One possibility is that a theory with particular flavor symmetries reduces the number of free parameters yielding precise relations between observables. Such a theory would lead to many zero matrix elements and/or to many precise relations between matrix elements. Here, we take a different, less ambitious approach. We aim to provide a general framework, Maximal Abelian Flavor Symmetries (MAFS), that regulates the sizes of all the matrix elements in such a way that the many small mass ratios and mixing angles can be derived, in an approximate way, from a few small parameters. This framework for a qualitative description of the flavor hierarchies has considerable generality, and may prove useful in constructing specific testable theories of flavor.

The idea of MAFS is that in any theory the magnitude of all flavor couplings is approximately determined by a set of real parameters  $\epsilon_a \lesssim 1$, one for every multiplet of Weyl fermions, $\psi_a$. For example, Yukawa couplings to scalars $\phi$ that contain the Higgs doublet are assumed to have the form
\begin{equation}
    {\cal L}_Y \; = \; C_{ab} \, \epsilon_a \epsilon_b \; (\psi_a \psi_b \,\phi) + \rm{h.c.}, \qquad |C_{ab}| = {\cal O}(1).
    \label{eq:MAFSinteraction}
\end{equation}
All the flavor hierarchies are described by $\epsilon_a$, which are fewer in number than the total number of couplings in the theory. For example, in the SM 15 $\epsilon_a$ parameters are used to estimate the 33 independent magnitudes in the flavor matrix elements. Crucial for further progress, the number of relevant $\epsilon_a$ parameters is significantly reduced in theories having fewer fermion multiplets; for example, there are 6 and 4 in the $SU(5)$ \cite{Georgi:1974sy} and $SO(10)$ \cite{Fritzsch:1974nn, Georgi:1974my} theories that we focus on in this paper.  We stress that MAFS alone is unable to make precise predictions due to the large number of order unity parameters in $C_{ab}$.

An early use of the MAFS idea was to control flavor-changing and CP-violating processes in multi-Higgs doublet extensions of the SM \cite{Antaramian:1992ya, Hall:1993ca}. A general theory with $N_H$ Higgs doublets has $54 N_H$ parameters in the flavor matrices of the charged fermions, and while many of these are not physical, those that are generically lead to unacceptably large rates for flavor and CP-violating processes unless the Higgs masses are very large, for example above $(10^3 - 10^4)$ TeV.  However, the number of fermions is unchanged, so with MAFS the fifteen $\epsilon_a$ parameters now determine hierarchies in all the flavor matrices. This results in flavor-violating processes allowing the extra Higgs scalars to be at the weak scale \cite{Antaramian:1992ya}. If the new CP phases are order unity, CP violation in the neutral kaons requires the extra scalars to be above the 10 TeV scale \cite{Hall:1993ca}.

MAFS is an alternative to the mechanism introduced by Froggatt and Nielsen \cite{Froggatt:1978nt}, where quark and charged lepton mass hierarchies arise from the seesaw mechanism by integrating out heavy fermions that are vector-like with respect to the SM gauge group. This mechanism prompted several studies of models in the following years \cite{1Dimopoulos:1983,2Berezhiani:1983,3Bagger:1984,4Davidson:1984,5Bagger:1985,6Berezhiani:1985,7Davidson:1988,8Davidson:1990,9Dimopoulos:1992,10Dimopoulos:1992b, Leurer:1992wg, Leurer:1993gy}. Most models of this widely-studied mechanism have a very small flavor group, just one or two $U(1)$s with one or two small symmetry breaking parameters,  $\epsilon$. Each fermion multiplet $\psi_a$, both SM and very heavy, has $U(1)$ charges $Q_a$, which take on small integer values. 
The flavor matrices in the full theory are typically, but not always, sparse, having many texture zeros.  With two $U(1)$s, for example, couplings $ \epsilon_1^{N_1}\epsilon_2^{N_2} \, \psi_a \psi_b \, \phi$ exist only if $|Q_{1a} + Q_{1b}| = N_1$ and $|Q_{2a} + Q_{2b}| = N_2$. 
For any given set of fermions there are many FN models, corresponding to the many choices for the integer charges; recent analyses use a Bayesian approach to scan over many thousands of models \cite{Ibe:2024cvi, Cornella:2023zme}. Applying MAFS to a set of fermions $\psi_a$, there is a single model described by $\epsilon_a$, which has no texture zeros. Both aim to provide an approximate understanding of the small dimensionless parameters in flavor physics. 

Many attempts to understand quark and lepton masses are based on non-Abelian flavor symmetries, both continuous \cite{Pomarol:1995xc, Barbieri:1995uv, King:2001uz} and discrete \cite{Hall:1995es, Ma:2001dn, Altarelli:2005yx}. They are very powerful and typically lead to texture zeros.  In this paper we take an alternative path, exploring the applications and implications of approximate Maximal Abelian Flavor Symmetries, or MAFS. In section \ref{sec:framework} we elaborate on the MAFS framework.
In section \ref{sec:321} we study MAFS with the SM gauge group, where the flavor symmetry is $U(1)^{15}$. While progress on understanding quark and lepton mass and mixing hierarchies is limited, MAFS provides a simple scheme for estimating the relative magnitudes of the many Wilson coefficients of SMEFT. We apply MAFS to theories with gauge group $G=SU(5)$ in section \ref{sec:5}; $U(1)^6$ allows significant progress. In section \ref{sec:10} we study MAFS with $G=SO(10)$; since the quarks and leptons are unified in a single spinor multiplet of $SO(10)$, it is not clear that MAFS can describe both small quark mixing and large neutrino mixing. We focus on the possibility that the SM Higgs lies in a spinor representation of $SO(10)$ \cite{Hall:2018let, Hall:2019qwx, Carrasco-Martinez:2025zus}, and then study more general possibilities. In section \ref{sec:lepto} we study whether this framework can successfully predict the order of magnitude of the cosmological baryon asymmetry generated via leptogenesis for these $G$.  In Appendix \ref{sec:texture0} we argue that texture zeros in a flavor matrix can be defined by an entry being much smaller than the reference value given by MAFS. In section \ref{sec:concl} we give a summary of our results for flavor hierarchies and leptogenesis for the three cases of $G = SU(3) \times SU(2) \times U(1), \;SU(5)$ and $SO(10)$.

\section{Framework}
\label{sec:framework}

We consider an effective field theory with a set of Weyl fermions, $\psi_a$, each in an irreducible representation of the gauge group $G$. 
For each $\psi_a$ there is an approximate Abelian symmetry which, for simplicity, we take to be $U(1)_a$. Hence, the flavor symmetry is the maximal Abelian symmetry $G_F = U(1)^{N_\psi}$, where $N_{\psi}$ is the number of $\psi_a$ fields. In the EFT, each $U(1)_a$ symmetry is explicitly broken by an amount described by a dimensionless parameter $\epsilon_a$, which is typically small but can be order unity. Taking $\psi_a$ to have unit charge under $U(1)_a$, the coupling of any interaction that carries $U(1)_a$ charge $Q_a$ is taken to be suppressed by a factor $(\epsilon_a)^{Q_a}$. Hence, any gauge-invariant Yukawa interaction between two fermions and a scalar $\phi$ is described by ${\cal L} \; = \; C_{ab} \, \epsilon_a \epsilon_b \; \psi_a \psi_b \, \phi + \rm{h.c.}$ with $|C_{ab}| = {\cal O}(1)$. Without loss of generality, $\epsilon_a$ are taken real, whereas $C_{ab}$ have phases, that we typically assume are order unity. For example, in a theory with the SM gauge group $\psi_a$ contains three generations of quark and lepton fields $(q_i, {\bar u}_i, \bar{d}_i, \ell_i, \bar{e}_i)$, leading to the MAFS group $G_F = U(1)^{15}$.

\begin{figure}[b]
\centering
\begin{tikzpicture}[>=stealth,thick,font=\small,
                    baseline=(current bounding box.center)]

\begin{scope}[shift={(0,0)}]
\draw[fermion,->] (-3,0) -- (-2.25,0);
\draw[fermion,-] (-2.25,0) -- (-1.5,0);
\draw[fermion,-] (-1.5,0) -- (-0.75,0);
\draw[fermion,<-] (-0.75,0) -- (0,0);

\draw[fermion,->] (0,0) -- (0.75,0);
\draw[fermion,-] (0.75,0) -- (1.5,0);
\draw[fermion,-] (1.5,0) -- (2.25,0);
\draw[fermion,<-] (2.25,0) -- (3,0);
\node[below] at (-3,0) {$\psi_a$};
\node[below] at ( 3,0) {$\psi_b$};
\node[below] at (0,0) {};
\node[below] at (-1.5,0) {};
\node[below] at (1.5,0) {};
\draw[scalar,-] (-1.5,0) -- (-1.5,1);
\draw[scalar,-] (-1.5,1) -- (-1.5,2);
\node[above] at (-1.5,2) {$\phi_{a}$};
\draw[scalar] (0,0) -- (0,2);
\node[above] at (0,2) {$\phi$};
\draw[scalar,-] (1.5,0) -- (1.5,1);
\draw[scalar,-] (1.5,1) -- (1.5,2);
\node[above] at (1.5,2)
  {$\phi_{b}$};

\node[below] at (-0.75,0)
  {$\Psi$};
\node[below] at (0.75,0)
  {$\Psi$};

\end{scope}
\end{tikzpicture}
\caption{UV completion with spontaneous breaking of $G_F$ by $\langle \phi_a \rangle$. This mixes $\psi_a$ with heavy vectorlike fermions $\Psi$ generating Yukawa couplings via the seesaw diagram.}
\label{fig:UVcompletion}
\end{figure}
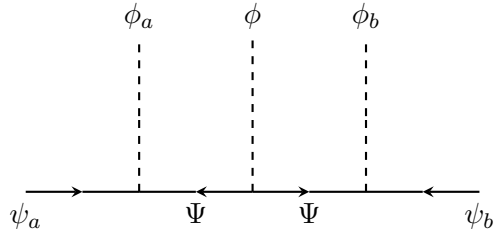

There are various possibilities for the UV origin of $G_F$ and the symmetry-breaking spurions $\epsilon_a$. For example, $G_F$ may be a UV symmetry spontaneously broken by scalar fields, $\phi_a$, so that all the Yukawa couplings of \ref{eq:MAFSinteraction} are generated via the seesaw diagram of Figure \ref{fig:UVcompletion}. The vectorlike fermions $\Psi$ have large masses $M_\Psi$ and are neutral under $G_F$. Gauge and flavor symmetries allow the Yukawa interactions $\psi_a \phi_a \Psi$ and $\Psi \Psi \phi$, which have order unity couplings; the scalars $\phi$ contain the SM Higgs doublet. Three generations of $\Psi$ are needed to give masses to the three generations in $\psi_a$. The MAFS symmetry breaking parameters are   $\epsilon_a \approx \langle \phi_a \rangle /M_\Psi$, some of which are typically smaller than the symmetry breaking parameter of the Froggatt-Nielsen mechanism. Unlike the Froggatt-Nielsen mechanism, there are no diagrams with long chains that generate operators beyond dimension 6. Alternatively, $G_F$ may arise as an accidental approximate IR symmetry from wavefunction overlaps in extra dimensions \cite{Arkani-Hamed:1999ylh}.

In this paper we focus on operators that lead to quark and lepton masses.  However, MAFS can be very easily applied to all operators in any Effective Field Theory (EFT). If physics at scale M generates an operator in the EFT of dimension 4+n, the MAFS estimate for its coefficient is a product of $\epsilon_a$, one for each $\psi_a$ in the operator, divided by $M^n$
\begin{equation}
    \mathcal{L}_{EFT} =  C'_{ab}\frac{\epsilon_a \epsilon_b}{M} \; \psi_a \,\sigma^{\mu \nu} F_{\mu \nu} \psi_b \; + \; C_{abcd} \frac{\epsilon_a \epsilon_b \epsilon_c \epsilon_d}{M^2} \; \psi_a \psi_b \; \psi_c \psi_d + ...
\label{eq:EFT}
\end{equation}
where $C'_{ab}, C_{abcd}$, like $C_{ab}$ of (\ref{eq:MAFSinteraction}), have order unity magnitudes and phases.
In SMEFT the number of operators grows rapidly with dimension, but MAFS allows a very simple estimate of all the coefficients using the $\epsilon_a$ values shown in Table \ref{tab:321eps}. This is similar in spirit to constraining SMEFT using maximal non-Abelian symmetries \cite{Faroughy:2020ina} or smaller non-Abelian groups \cite{Greljo:2025mwj}.

Any flavor observable, CP conserving or not, can be estimated using the ``MAFS approximation'' by simply including the relevant factors of $\epsilon_a$. The uncertainty in the estimate depends on the distribution for the $C$ coefficients about unity, and on how many such coefficients appear in the expression for the observable.

In theories with MAFS, the set of fermion multiplets, $\psi_a$, may include ones with very heavy states, $\psi_A$, as well as those that contain SM states, $\psi_\alpha$. Seesaw contributions to SM fermion masses arise when $\psi_A$ are integrated out. However, in the seesaw diagram factors of $\epsilon_A$ cancel between numerator and denominator, so that the resulting EFT of the SM has flavor interactions that only involve $\epsilon_\alpha$. An exception to this can result if the heavy fermion masses arise at a stage of gauge symmetry breaking such that the identity of heavy and light states is interchanged, as we will discover in the $SO(10)$ model of section \ref{sec:10}.

\section{$G = SU(3) \times SU(2) \times U(1)$}
\label{sec:321}

Consider theories with gauge group $G = SU(3) \times SU(2) \times U(1)$ and minimal fermions;  three generations of quark and lepton fields $\psi_a =(q_i, {\bar u}_i, \bar{d}_i, \ell_i, \bar{e}_i)$, leading to the MAFS group $G_F = U(1)^{15}$. The scalar sector of the theory can be arbitrary; for example, containing more than one electroweak Higgs doublet. We define $h$ to be the combination of doublets that acquires a vev, so that the Yukawa interactions for charged fermion masses and the dimension-5 operators for neutrino masses are
\begin{equation}
    \mathcal{L}_{SM, F} =  Y^U_{ij} \, q_i {\bar u}_j \, h +
    Y^D_{ij} \, q_i {\bar d}_j \, h^\dagger +
    Y^E _{ij} \, \ell_i {\bar e}_j \, h^\dagger + Y^{(5)}_{ij} \, \ell_i \ell_j \; \frac{hh}{M} \; + \rm{h.c.}
\label{eq:L321}
\end{equation}
with
\begin{equation}
       Y^U_{ij} \; = \; C^U_{ij} \, \epsilon^q_i \epsilon^{\bar u}_j,\hspace{0.2in}
       Y^D_{ij} \;=\; C^D_{ij} \, \epsilon^q_i \epsilon^{\bar d}_j,\hspace{0.2in}
       Y^E_{ij} \;=\; C^E_{ij} \, \epsilon^\ell_i \epsilon^{\bar e}_j,\hspace{0.2in}
       Y^{(5)}_{ij} = C^{(5)}_{ij} \, \epsilon^\ell_i \epsilon^\ell_j.
\label{eq:Y321}
\end{equation}
$C^{U,D,E,{(5)}}_{ij}$ are all near unity and $M$ is a large mass scale associated with lepton number violation. The phases in $C_{ij}$ are the origin of CP violation, accounting for the order unity phase in the CKM matrix and leading to the expectation of order unity phases in the PMNS leptonic mixing matrix. Other Higgs doublets have very similar interactions, with alternative order unity coefficients, that could give interesting flavor-changing and CP-violating signals if they have TeV-scale masses \cite{Antaramian:1992ya, Hall:1993ca}. The flavor structure of higher-dimensional operators is also governed by $\epsilon^{q, {\bar u}, \bar{d}, \ell, \bar{e}}_i$.

The masses and mixings of the quarks and leptons can be described by the 15 symmetry-breaking parameters $\epsilon^{q, {\bar u}, \bar{d}, \ell, \bar{e}}_i$, with all $|C_{ij}|$ within a factor 2 of unity. We take $\epsilon^x_1 < \epsilon^x_2 < \epsilon^x_3$, $x= q, {\bar u}, \bar{d}, \ell, \bar{e}$. All mass and mixing angle hierarchies are taken to arise from the small parameters in $\epsilon^{q, {\bar u}, \bar{d}, \ell, \bar{e}}_i$. For example, $\det Y^E$ is the sum of 6 terms, all proportional to  $(\epsilon^\ell_1 \epsilon^{\bar e}_1) (\epsilon^\ell_2 \epsilon^{\bar e}_2) (\epsilon^\ell_3 \epsilon^{\bar e}_3)$ with each having independent coefficients trilinear in $C^E_{ij}$. We do not allow fine-tuning between these six terms so we expect the eigenvalues $y_{(e,\mu,\tau)}$ to be of order $(\epsilon^\ell_1 \epsilon^{\bar e}_1, \, \epsilon^\ell_2 \epsilon^{\bar e}_2, \, \epsilon^\ell_3 \epsilon^{\bar e}_3)$.

It is important to estimate $\epsilon^{q, {\bar u}, \bar{d}, \ell, \bar{e}}_i$ from the data. These estimates have uncertainties due to the unknown order unity parameters $C_{ij}$. We cannot get an estimate by simply setting $C^{U,D,E,(5)}_{ij} = 1$, because in this limit the masses of the quarks and leptons of the lightest two generations are zero and there is no flavor mixing.  Instead, we assume values for $C^{U,D,E,(5)}_{ij}$ that set the eigenvalues of the Yukawa coupling and neutrino mass matrices to approximate values determined solely by the dependence on $\epsilon^{q, {\bar u}, \bar{d}, \ell, \bar{e}}_i$:
\begin{equation}
       y_{u_i} \; \approx \epsilon^q_i \epsilon^{\bar u}_i,\hspace{0.2in}
       y_{d_i} \; \approx \epsilon^q_i \epsilon^{\bar d}_i,\hspace{0.2in}
       y_{e_i} \; \approx \epsilon^\ell_i \epsilon^{\bar e}_i,\hspace{0.2in}
       m_{\nu_i} \; \approx \epsilon^\ell_i \epsilon^\ell_i \; {v^2/M},
       \label{eq:321masses}
\end{equation}
where $v=174$ GeV is the vev of the Higgs field.
Similarly, the elements of the CKM and PMNS mixing matrices are taken to be
\begin{equation}
 |V_{ij}| \approx \epsilon^{q}_i / \epsilon^{q}_j,\hspace{0.3in} |U_{ij}| \approx \epsilon^{\ell}_i / \epsilon^{\ell}_j, \qquad j>i.
 \label{eq:321mixings}
\end{equation}
Results that show only the dependence of observables on the symmetry breaking parameters, such as (\ref{eq:321masses}) and (\ref{eq:321mixings}), we call predictions in the MAFS approximation.

In the quark sector there are 9 $\epsilon$ parameters for 9 observables. However, because the CKM matrix elements of (\ref{eq:321mixings}) depend only on ratios of $\epsilon^q_i$, there are predictions\footnote{Here, and elsewhere in the paper, equations with an approximate sign indicate results that follow from assuming the MAFS approximation.}
\begin{equation}
V_{ub} \; \approx \; V_{td} \; \approx \; V_{us} V_{cb}.
\label{eq:321VCKM}
\end{equation}
which are correct at the factor of two level. The three sides of the CKM unitarity triangle are predicted to have comparable lengths, so that the three angles ($\alpha, \beta, \gamma$) are all order unity. 
We determine $\epsilon^{q}_2 / \epsilon^{q}_3$ from $V_{cb}$, and then determine $\epsilon^{q}_1 / \epsilon^{q}_2$ from $V_{us}$ {\em or} $\,\epsilon^{q}_1 / \epsilon^{q}_3$ from $V_{ub}$. We choose the latter, where the value of $\epsilon_1$ relative to $\epsilon_{2,3}$ is smaller by a factor 2, as it improves consistency with $SU(5)$ unification, discussed in the next section. 
This leaves 7 $\epsilon$ parameters for 6 quark masses. However, since the top Yukawa coupling is near unity, we must take both $\epsilon^q_3$ and $\epsilon^{\bar{u}}_3$ to be order unity; anticipating that $SU(5)$ unification works well for the third generation, we take them equal. We then determine $\epsilon^{\bar{u}}_{1,2}$ from $y_{u,c}$ and $\epsilon^{\bar{d}}_{1,2,3}$ from $y_{d,s,b}$. 

In the lepton sector, the large mixing angles imply, from (\ref{eq:321mixings}), that the $\epsilon^\ell_i$ are not as hierarchical as the $\epsilon^q_i$; indeed, they may not have any hierarchy. The Abelian lepton symmetries do not impose approximate degeneracies on the neutrinos, so that the inverted and quasi-degenerate neutrino spectra are highly disfavored; the neutrino spectrum has normal ordering. If there is no hierarchy between $\epsilon^\ell_2$ and $\epsilon^\ell_3$,  $m_{\nu_2}/m_{\nu_3} \simeq 0.2$ must arise entirely from deviations of $C^{(5)}_{ij}$ from unity. We prefer to impose a mild hierarchy of $\epsilon^\ell_2/\epsilon^\ell_3 \approx 1/2$ to account for this mass ratio. Similarly, the measured values of $U_{ij}$ suggest a mild hierarchy between $\epsilon^\ell_1$ and $\epsilon^\ell_3$, and we adopt $\epsilon^\ell_1/\epsilon^\ell_3 \approx 1/4$. In this case, we predict a substantial mass for $\nu_1$ and also for the $ee$ entry of the neutrino mass matrix, relevant for neutrinoless double beta decay
\begin{equation}
m_{\nu_1} \;\approx \; 0.003 \,\mbox{eV}, \hspace{1in}
m_{\nu_{ee}} \;\approx \; 0.003 \,\mbox{eV}. 
\label{eq:mnu1}
\end{equation}
Next generation neutrinoless double beta decay experiments aim to probe $m_{\nu_{ee}}$ of 0.01 eV or below \cite{Dolinski:2019nrj,nEXO:2021ujk,LEGEND:2021bnm}. We stress that these MAFS predictions are not precise, having uncertainties of factors of 2.

\begin{figure}[]
\hspace{-0.6cm}
\begin{tikzpicture}[x=1.82cm,y=1.25cm,font=\normalsize]
\tikzset{
  Eone/.style={draw=blue!50!black, fill=blue!90!black, circle, minimum size=6pt, inner sep=1pt},
  Etwo/.style={draw=green!25!black, fill=green!65!black, rectangle, minimum size=6pt, inner sep=1pt},
  Ethree/.style={draw=red!37!black, fill=red!77!black, regular polygon,
    regular polygon sides=3, minimum size=8pt, inner sep=1pt},
  numlabel/.style={
    font=\footnotesize,
    anchor=west,
    fill=white,
    fill opacity=0.90,
    text opacity=1,
    inner sep=1.0pt,
    rounded corners=1pt
  }
}
\draw[->] (-0.5,-6.5) -- (-0.5,0.25);
\node[rotate=90,font=\large] at (-1.12,-3.15) {$\log_{10}(y)$};
\foreach \y in {0,-1,-2,-3,-4,-5,-6} {
  \draw[dashed, gray!35] (-0.5,\y) -- (7.05,\y);
  \node[anchor=east,font=\small] at (-0.58,\y) {$\y$};
}
\def\xUp{0.35}
\def\xDown{2.55}
\def\xLept{4.65}
\node[align=center,font=\normalsize] at (\xUp+0.25,0.85) {Up-type};
\node[align=center,font=\normalsize] at (\xDown+0.25,0.85) {Down-type};
\node[align=center,font=\normalsize] at (\xLept+0.25,0.85) {Leptons};

\foreach \x/\pname/\lA/\vA/\lB/\vB/\lC/\vC in {
 \xUp/t/{-0.15}/{0.71}/{-0.31}/{0.49}/{-0.373}/{0.42},
 \xUp/c/{-2.60}/{2.5{\times}10^{-3}}/{-2.80}/{1.6{\times}10^{-3}}/{-2.864}/{1.4{\times}10^{-3}},
 \xUp/u/{-5.31}/{4.9{\times}10^{-6}}/{-5.49}/{3.2{\times}10^{-6}}/{-5.563}/{2.7{\times}10^{-6}},
 \xDown/b/{-1.96}/{1.1{\times}10^{-2}}/{-2.16}/{6.8{\times}10^{-3}}/{-2.239}/{5.8{\times}10^{-3}},
 \xDown/s/{-3.67}/{2.2{\times}10^{-4}}/{-3.85}/{1.4{\times}10^{-4}}/{-3.915}/{1.2{\times}10^{-4}},    
 \xDown/d/{-4.97}/{1.1{\times}10^{-5}}/{-5.15}/{7.1{\times}10^{-6}}/{-5.215}/{6.1{\times}10^{-6}},
 \xLept/\tau/{-1.99}/{1.0{\times}10^{-2}}/{-1.99}/{9.9{\times}10^{-3}}/{-2.015}/{9.7{\times}10^{-3}},
 \xLept/\mu/{-3.22}/{6.0{\times}10^{-4}}/{-3.24}/{5.8{\times}10^{-4}}/{-3.248}/{5.6{\times}10^{-4}},
 \xLept/e/{-5.54}/{2.9{\times}10^{-6}}/{-5.56}/{2.8{\times}10^{-6}}/{-5.567}/{2.7{\times}10^{-6}}
}{
  \draw[black!80!black, thick]
  (\x-0.30,\lA) -- (\x,\lB) -- (\x+0.30,\lC);

  \node[Eone] at (\x-0.30,  \lA) {};
  \node[Etwo] at (\x ,     \lB) {};
  \node[Ethree] at (\x+0.30,   \lC) {};

\node[anchor=south,font=\small] at (\x,\lB+0.11) {$\pname$};

  \node[numlabel,font=\small] at (\x+0.42,\lB+0.0)
  {\begin{tabular}{@{}c@{}}
  $\vA$\\[0pt]
  $\vB$\\[0pt]
  $\vC$\\[0pt]
  \end{tabular}};
}

\begin{scope}[shift={(6.1,0.83)}]
  \node[Eone] at (0,0) {};
  \node[anchor=west,font=\small] at (0.05,0) {$10^{5}\,\mathrm{GeV}$};

  \node[Etwo] at (0,-0.35) {};
  \node[anchor=west,font=\small] at (0.05,-0.35) {$10^{12}\,\mathrm{GeV}$};

  \node[Ethree] at (0,-0.70) {};
  \node[anchor=west,font=\small] at (0.05,-0.70) {$10^{16}\,\mathrm{GeV}$};
\end{scope}

\end{tikzpicture}
\caption{Yukawa couplings for up-type quarks, down-type quarks, and charged leptons evaluated at renormalization scales $\mu = 10^{5},\,10^{12},\,10^{16}\,\mathrm{GeV}$, based on~\cite{Huang:2020hdv}.
}
\label{fig:PanelofYukawas}
\end{figure}
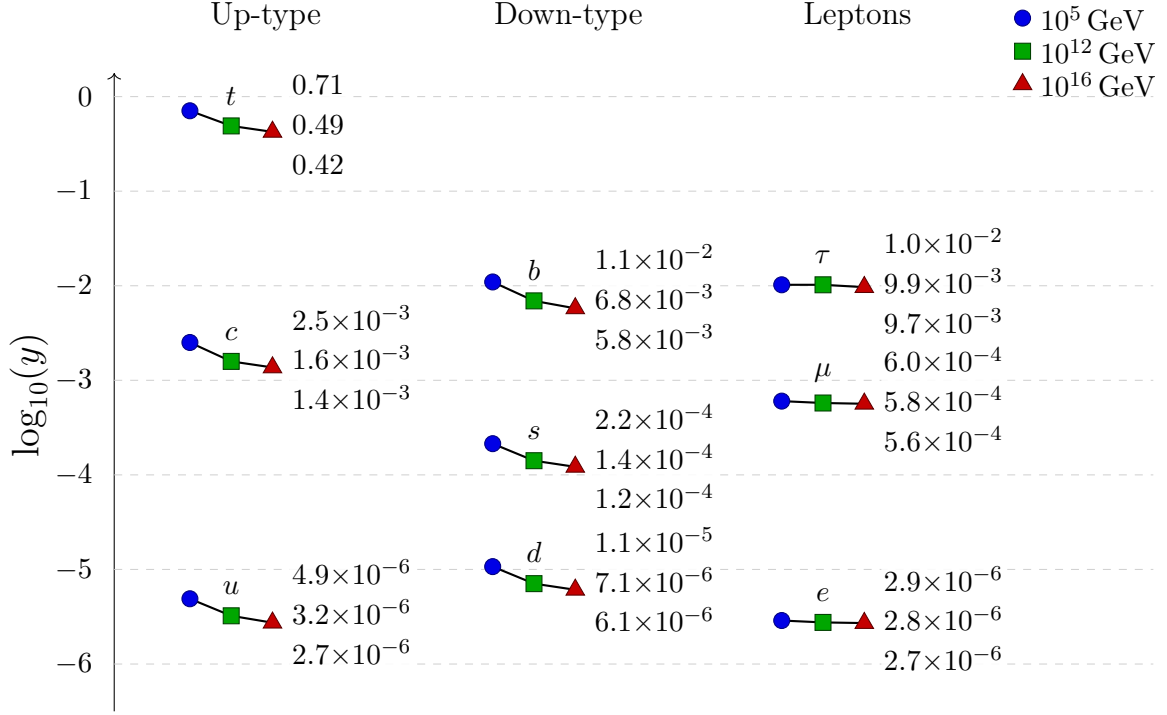

Neutrino masses and mixings leave $\epsilon^\ell_3$ and $\epsilon^{\bar{e}}_i$ undetermined. The three charged lepton masses fix $\epsilon^{\bar{e}}_i$ in terms of $\epsilon^\ell_3$.  Our estimations for $\epsilon^{q, {\bar u}, \bar{d}, \ell, \bar{e}}_i$ in the $\epsilon$-scaling approximation are shown in Table \ref{tab:321eps}, where we have used the Yukawa couplings of Fig. \ref{fig:PanelofYukawas}, and quark and lepton mixing angles from PDG. Finally, the mass scale $M$ is determined from $m_{\nu_3}$ in terms of $\epsilon^\ell_3$
\begin{equation}
M \;\approx \; 6 \times 10^{14} \, (\epsilon^\ell_3)^2 \,\mbox{GeV}.
\label{eq:321M}
\end{equation}
Since $\epsilon^\ell_3$ and $\epsilon^{\bar{e}}_3$ are both of order unity or less, we deduce that $10^{-2} \lesssim \epsilon^\ell_3 \lesssim 1$, so that $6 \times 10^{10} \, \mbox{GeV} \, \lesssim M \, \lesssim 6 \times 10^{14} \, \mbox{GeV}$. Hence, this mass scale of lepton number violation is predicted to be below the scale of grand unification, but large enough that decays of right-handed neutrinos could yield the cosmological baryon asymmetry by thermal leptogenesis. 

To obtain the numerical results of Table \ref{tab:321eps}, one must make a choice for the renormalization scale at which (\ref{eq:321masses}) and (\ref{eq:321mixings}) apply. This is not important for lepton Yukawas and mixings, since the running is so mild. Even the running from quark Yukawa couplings, shown in Fig. \ref{fig:PanelofYukawas},  may not be larger than the uncertainties arising from the $C_{ij}$; still, for concreteness, we evaluate the SM Yukawa couplings at a scale of $10^{16}$ GeV, relevant when the flavor symmetries are defined at the unification scale. The above procedure then leads to the values of $\epsilon^{q, {\bar u}, \bar{d}, \ell, \bar{e}}_i$ shown in Table \ref{tab:321eps}. If matching is done at the TeV scale, then $\epsilon^{q, {\bar u}, \bar{d}}_i$ should be increased by about 50\%. 
We stress that the uncertainties from the unknown parameters $C_{ij}$ are important and therefore we only quote the $\epsilon$ values to one significant figure. It may be thought that a factor 2 uncertainty in $C_{ij}$ only leads to factors of $\sqrt{2}$ uncertainties in the $\epsilon$ when fitting to quark and lepton masses. While this is true for 3rd generation masses, second (first) generation masses have 2 (6) contributions with arbitrary signs, increasing the uncertainties. Similarly, mixing angles have contributions from two sectors (u/d or e/$\nu$), which can have equal or opposite signs.

\begin{table}[]
\centering
\setlength{\tabcolsep}{12pt}
\begin{tabular}{|c|c|c|c|}
\hline
 & $\epsilon_1$ & $\epsilon_2$ & $\epsilon_3$ \\
\hline
$q$ & $\approx 3 \times 10^{-3}$ & $\approx 3 \times 10^{-2}$ & $\approx 0.7$ \\
$\bar{u}$ & $\approx 1.2 \times 10^{-3}$ & $\approx 6 \times 10^{-2}$ & $\approx 0.7$ \\
$\bar{d}$ & $\approx 2.6 \times 10^{-3}$ & $\approx 5 \times 10^{-3}$ & $\approx 1.0 \times 10^{-2}$ \\
$ \ell $ & $\approx 0.25 \, \epsilon^{\ell}_3$ & $\approx 0.5 \, \epsilon^{\ell}_3$ &  $\epsilon^{\ell}_3$ \\
$\bar{e}$ & $\approx 1.1 \times 10^{-5} / \epsilon^\ell_3$ & $\approx 1.2 \times 10^{-3} / \epsilon^\ell_3$ & $\approx 1.0 \times 10^{-2} / \epsilon^\ell_3$ \\
\hline
\end{tabular}
\caption{Values of $\epsilon^{q, {\bar u}, \bar{d}, \ell, \bar{e}}_i$ for $G=SU(3) \times SU(2) \times U(1)$ extracted from data using approximations described in the text.}
\label{tab:321eps}
\end{table}

\section{$SU(5)$ Unification}
\label{sec:5}
MAFS become more powerful and predictive as gauge unification reduces the number of independent multiplets in a generation. With $G = SU(5)$, the 15-plet generation fits into $\bf{\bar{5}} + \bf{10}$, with fields $\bar{F}(\bar{d}, \ell), T(q, \bar{u}, \bar{e})$. For each generation we add an $SU(5)$ singlet field $\bar{\nu}$. These states are lighter than $V_5$, the mass scale of $SU(5)$ breaking, and are responsible for generating the observed neutrino masses. The SM Higgs doublet must lie in scalar fields that transform in representations formed from the product of two of these fermions, i.e. in $\bf{5}$ or $\bf{45}$.  Without loss of generality, we take the SM doublet to lie partly in $H_1({\bf 5})$ and partly in $H_2({\bf 45})$. Since no flavor symmetry acts on the scalars, the two components are comparable;
flavor hierarchies do not arise from the scalar sector. The most general set of Yukawa interactions that generate the charged fermion masses is 
\begin{equation}
    \mathcal{L}_{SU(5), Y} \; = \; {\cal Y}^{TT}_{ijA} \; T_i T_j \, H_A \; + \;
    {\cal Y}^{TF}_{ijA} \; T_i {\bar F}_j \, H_A^\dagger 
     \; + \rm{h.c.}, \qquad A=1,2.
\label{eq:L5Y}
\end{equation}

We do not include higher-dimension operators that give contributions to the fermion masses suppressed by powers of $V_5/\Lambda$, where $\Lambda$ is the cutoff scale of the $SU(5)$ EFT. Even though $\Lambda$ need not be far above $V_5$, we assume such operators are subject to the same flavor symmetry suppression as the Yukawa couplings, and hence these operators are expected to make subdominant contributions to flavor observables. The theory also contains Yukawa interactions of the form $\bar{F}_i \bar{\nu}_j H_1$, and integrating out the $\bar{\nu}$ states leads to the dimension 5 operator that generates neutrino masses
\begin{equation}
    \mathcal{L}_\nu \; = \;  Y^{(5)}_{ij} \; \ell_i\ell_j \, \frac{hh}{M}
     \; + \rm{h.c.}
\label{eq:L5nu5}
\end{equation}
where $h$ is the SM Higgs. Unlike (\ref{eq:L5Y}), this operator is not $SU(5)$ invariant because $M \ll V_5$ and the color-triplet partners of the SM Higgs have mass of order $V_5$.

It is well-known that if $H_2({\bf 45})$ is absent, so that $h$ is contained entirely in $H_1(\bf{5})$, the second interaction of (\ref{eq:L5Y}) leads to the predictions $y_{d_i} = y_{e_i}$. Running the observed SM Yukawa couplings to the scale $10^{16}$ GeV, gives $y_b/y_\tau = 0.68, \, y_s/y_\mu = 0.22$ and $y_d/y_e = 2.3$. Clearly, there must be large breaking of these $SU(5)$ relations in the first two generations and it is for this reason that we insist that $h$ has an order unity component in $H_2({\bf 45})$.\footnote{Georgi and Jarlskog introduced a scheme for obtaining $y_b/y_\tau = 1, \, y_s/y_\mu = 1/3$ and $y_d/y_e = 3$ at the unification scale \cite{Georgi:1979df}. However, they require a very particular form for the flavor symmetries that must also act on the scalar multiplets. Thus their model lies outside the general set of theories we have constructed. Furthermore, each of their relations requires order 30\% corrections, for example from higher-dimensional operators.} 
While the component of h in $H_1({\bf 5})$ gives equal contributions to charged lepton and down quark masses, that in $H_2({\bf 45})$ leads to a contribution that is 3 times larger for the charged leptons than for the down quarks. Hence, as we will see, with approximate flavor symmetries the combined contributions from both $H_1$ and $H_2$ give approximate mass relations $y_{d_i} \approx y_{e_i}$.

The MAFS relevant for the SM fermion masses is $G_F = U(1)^6$, with breaking characterized by the 6 parameters $\epsilon^T_i$ and $\epsilon^{\bar{F}}_i$, giving
\begin{equation}
       {\cal Y}^{TT}_{ijA} = C^{TT}_{ijA} \;\epsilon^T_i \epsilon^T_j,\hspace{0.2in}
       {\cal Y}^{TF}_{ijA}  \;=\; C^{TF}_{ijA} \; \epsilon^T_i \epsilon^{\bar{F}}_j,\hspace{0.2in}
       Y^{(5)}_{ij} = C^{(5)}_{ij} \, \epsilon^{\bar{F}}_i \epsilon^{\bar{F}}_j.
\label{eq:calY5}
\end{equation}
The SM Yukawa couplings of (\ref{eq:L321}) are
\begin{equation}
       Y^U_{ij} \; = \; C^U_{ij} \, \epsilon^T_i \epsilon^T_j,\hspace{0.2in}
       Y^D_{ij} \;=\; C^D_{ij} \, \epsilon^T_i \epsilon^{\bar{F}}_j,\hspace{0.2in}
       Y^E_{ij} \;=\; C^E_{ij} \, \epsilon^{\bar{F}}_i\epsilon^T_j ,\hspace{0.2in}
       Y^{(5)}_{ij} = C^{(5)}_{ij} \, \epsilon^{\bar{F}}_i \epsilon^{\bar{F}}_j.
\label{eq:Y5}
\end{equation}
The $C^{U,D,E, (5)}_{ij}$ are again all near unity, but in general take different values from the non-unified case with $G = SU(3) \times SU(2) \times U(1)$. In particular, if $h$ has a component $\cos \theta_h$ in $H_1$ and $\sin \theta_h$ in $H_2$, then 
\begin{equation}
       C^D_{ij} \;=\; \cos \theta_h \, C^{TF}_{ij1} - \frac{1}{3}\sin \theta_h \, C^{TF}_{ij2},\hspace{0.2in}
       C^E_{ij} \;=\; \cos \theta_h \, C^{TF}_{ij1} + \sin \theta_h \, C^{TF}_{ij2}.
\label{eq:CDE}
\end{equation}
$C^D_{ij}$ and $C^E_{ij}$ differ due to contributions from the component of $h$ in $H_2({\bf 45})$; hence, the d/e mass relations are approximate and not exact.

\begin{table}[]
\centering
\setlength{\tabcolsep}{12pt}
\begin{tabular}{|c|c|c|c|}
\hline
 & $\epsilon_1$ & $\epsilon_2$ & $\epsilon_3$ \\
\hline
$q$ & $\approx 3 \times 10^{-3}$ & $\approx 3 \times 10^{-2}$ & $\approx 0.7$ \\
$\bar{u}$ & $\approx 1.2 \times 10^{-3}$ & $\approx 6 \times 10^{-2}$ & $\approx 0.7$ \\
$\bar{e}$ & $\approx 1.1 \times 10^{-3}$ & $\approx 12 \times 10^{-2} $ & $\approx 1.0 $ \\
\hline
$\bar{d}$ & $\approx 2.6 \times 10^{-3}$ & $\approx 5 \times 10^{-3}$ & $\approx 1.0 \times 10^{-2}$ \\
$ \ell $ & $\approx 2.5 \times 10^{-3}$ & $\approx 5 \times 10^{-3}$ &  $\approx 1.0 \times 10^{-2}$ \\
\hline
\end{tabular}
\caption{Values of $\epsilon^{q, {\bar u}, \bar{d}, \ell, \bar{e}}_i$ for $G=SU(3) \times SU(2) \times U(1)$ extracted from data using approximations described in the text, with $\epsilon^\ell_3$ set to $10^{-2}$.}
\label{tab:321-5eps}
\end{table}
Are the observed quark and lepton masses and mixings consistent with (\ref{eq:Y5})? This is far from obvious, given that there are only six independent $\epsilon^{T, \bar{F}}_i$ parameters. Since $T = (q, \bar{u}, \bar{e})$ and $\bar{F} = (\bar{d}, \ell)$, the $SU(5)$ gauge symmetry imposes $\epsilon^q_i = \epsilon^{\bar u}_i = \epsilon^{\bar e}_i$ and $\epsilon^{\bar d}_i = \epsilon^\ell_i$. We can gain insight by investigating whether our results for $G = SU(3) \times SU(2) \times U(1)$ shown in Table \ref{tab:321eps} are consistent with this. Choosing $\epsilon^\ell_3 = 10^{-2}$ yields the numbers shown in Table \ref{tab:321-5eps}. We see there is remarkable consistency with $\epsilon^q_3 = \epsilon^{\bar u}_3 = \epsilon^{\bar e}_3$ and $\epsilon^{\bar d}_i = \epsilon^\ell_i$, while the differences within the $\epsilon^{q, {\bar u}, \bar{e}}_1$ and within the $\epsilon^{q, {\bar u}, \bar{e}}_2$ are only factors of 2 and 3. Next, we confirm this success by extracting $\epsilon^{T, \bar{F}}_i$ directly from data.    

In the MAFS approximation, 18 observables are described by 7 parameters (6 $\epsilon$ parameters and $M$). The 11 approximate predictions are 
\begin{equation}
       y_{d_i} \approx y_{e_i}, \hspace{0.4in}  V_{ij}  \approx  \sqrt{\frac{y_{u_i}}{y_{u_j}}}, \hspace{0.4in}  \sqrt{\frac{m_{\nu_i}}{m_{\nu_j}}} \approx U_{ij}  \approx  \frac{y_{d_i}}{y_{d_j}} \sqrt{\frac{y_{u_j}}{y_{u_i}}},  \hspace{0.4in} j>i.
\label{eq:11pred}
\end{equation}
This implies that, when we use the MAFS approximation to extract $\epsilon^{\bar{F}}_i$, the result will depend on whether we use $y_{d_i}$ or $y_{e_i}$.  In an attempt to obtain optimal values for $\epsilon^{\bar{F}}_i$ we use the geometric means, $\sqrt{y_{d_i} y_{e_i}}$. Thus we choose to determine $\epsilon^{T,\bar{F}}_i$ from the charged fermion masses via 
\begin{equation}
       (\epsilon^T_i)^2 \approx y_{u_i}, \hspace{0.4in} 
       \epsilon^T_i \epsilon^{\bar{F}}_i\approx \sqrt{y_{d_i} y_{e_i}},
\label{eq:epsTF}
\end{equation}
giving the results in Table~\ref{tab:5epsTF}.
\begin{table}[b]
\centering
\renewcommand{\arraystretch}{1.1}
\setlength{\tabcolsep}{12pt}
\begin{tabular}{|c|c|c|c|}
\hline
 & $\epsilon_1$ & $\epsilon_2$ & $\epsilon_3$ \\
\hline
$T$ & $\approx 1.8\times10^{-3}$ & $\approx 4.0\times 10^{-2}$ & $\approx0.65$ \\
$\bar{F}$ & $\approx 2.3\times10^{-3}$ & $\approx 6.6\times10^{-3}$ & $\approx 1.3\times10^{-2}$ \\
\hline
\end{tabular}
\caption{Values of $\epsilon^{T,\bar{F}}_i$ in the $SU(5)$ theory extracted using the MAFS approximation of (\ref{eq:epsTF}).}
\label{tab:5epsTF}
\end{table}
The key point is that, while the breaking of $U(1)^3$ on $T_i$ is hierarchical, with $\epsilon^T_1 \ll \epsilon^T_2 \ll \epsilon^T_3$, there is almost no hierarchy in the breaking of $U(1)^3$ on $\bar{F}_i$, since $\epsilon^{\bar{F}}_1 \sim (1/3) \epsilon^{\bar{F}}_2$ and $\epsilon^{\bar{F}}_2 \sim (1/2) \epsilon^{\bar{F}}_3$. This then guarantees that while the CKM mixing angles on the quarks are predicted to be small
\begin{equation}
       V_{ij} \approx \frac{\epsilon^T_i}{\epsilon^T_j} \ll 1 \hspace{0.4in} i<j,
\label{eq:5hier}
\end{equation}
the leptonic mixing angles and neutrino mass ratios are predicted to have small hierarchies
\begin{equation}
       U_{ij} \; \approx \; \frac{\epsilon^{\bar{F}}_i}{\epsilon^{\bar{F}}_j},  \hspace{0.4in} \frac{m_{\nu_i}}{m_{\nu_j}} \; \approx \; \left( \frac{\epsilon^{\bar{F}}_i}{\epsilon^{\bar{F}}_j} \right)^2,  \hspace{0.4in}
       i<j.
\label{eq:5nohier}
\end{equation}
Indeed, using MAFS in simple $SU(5)$ unified theories, large leptonic mixing angles and order unity neutrino mass ratios are predicted from the observed quark and charged lepton masses. This arises because the hierarchy in masses in the up sector is roughly the square of that in the down quark and charged lepton sectors. This understanding of large neutrino masses from MAFS is different from previous ideas for generating large neutrino mixing angles in unified theories without a flavor symmetry \cite{Babu:1995uu}, with a single $U(1)$ flavor symmetry \cite{Altarelli:1998ns, Yanagida:1998jk, Buchmuller:1998zf}, 
or with {\em all} the small parameters of the flavor sector arising from the mixing of $T_i, \bar{F}_i$ with additional fermion multiplets \cite{Babu_1996}. For early models with large neutrino mixing angles resulting from a $U(2)$ flavor symmetry in unified theories see \cite{Carone:1997qg, Barbieri:1999pe}. More recently, it has been commented that neutrino anarchy is consistent with quark hierarchies in $SU(5)$ if $U(2)$ acts on $T_{1,2}$ but not on $\bar{F}_{1,2}$ \cite{Antusch:2023shi}.

The lightest neutrino mass and the mass relevant for neutrinoless double beta decay are predicted to be $m_{\nu_1}, m_{ee} \approx (1-2) 10^{-3}$ eV. Using $\epsilon^{\bar{F}}_{2,3}$ of Table~\ref{tab:5epsTF}, the normalization of the observed neutrino masses requires the mass scale of the dimension-5 operator of (\ref{eq:L321}) to be
\begin{equation}
M \;\approx \; 10^{11} \,\mbox{GeV}
\label{eq:5M}
\end{equation}
which is consistent with setting $\epsilon^\ell_3 = 10^{-2}$ in (\ref{eq:321M}).

\begin{figure}[]
    \centering
    \includegraphics[width=1.0\linewidth]{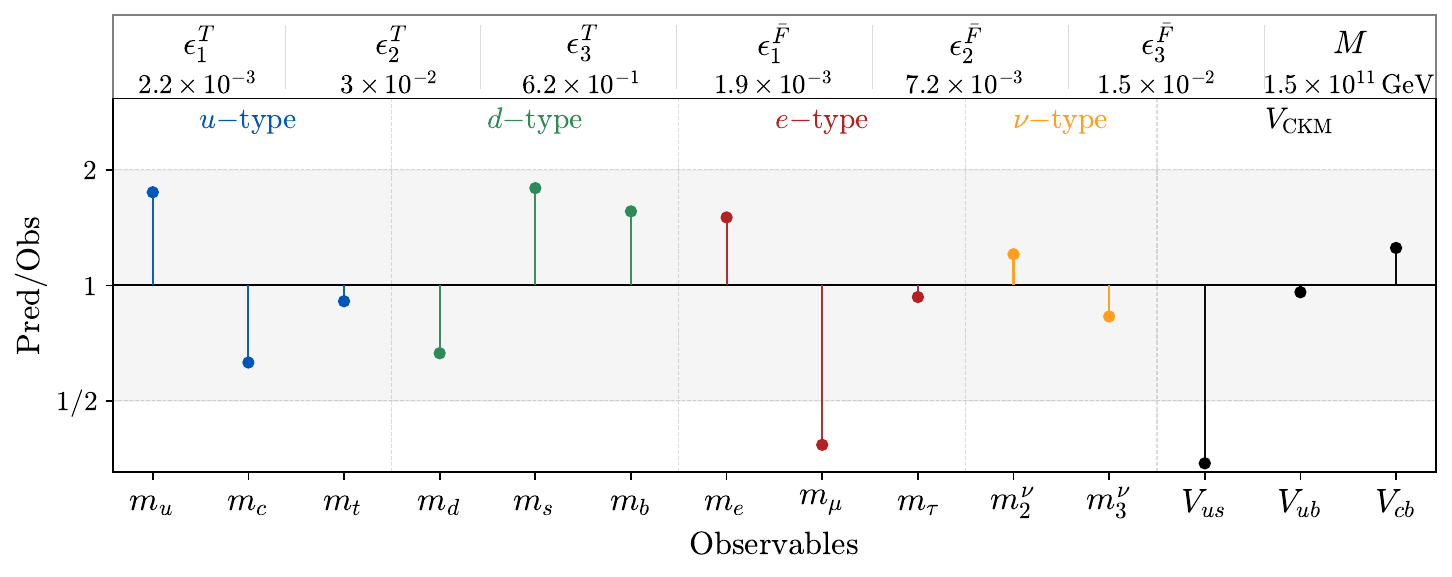}
    \caption{Results of a fit in the $SU(5)$ theory to 14 observables (quark and lepton masses and CKM mixings) with seven parameters  ($\epsilon^{T, \bar{F}}_i, M$) using the MAFS approximation.}
    \label{fig:SU5fit}
\end{figure}

These results are confirmed by fitting 14 observables $O_I$ (quark and lepton masses and CKM mixings) to the seven free parameters, $p_a$ ($\epsilon^{T, \bar{F}}_i, M$),  by minimizing  $d
^2(p_a) = \sum_I\ln^2(O_I/P_I(p_a))$ with respect to $p_a$, where $P_I(p_a)$ is the prediction for observable $I$. The results of the fit are shown in Fig. \ref{fig:SU5fit}. 
The parameters take values close to those given in Table \ref{tab:5epsTF}.
All 14 observables take predicted values that agree with data to better than a factor 2, except for $m_\mu$ and $V_{us}$, which are too small by about a factor of 3. We conclude that the MAFS approximation works remarkably well in $SU(5)$. While contributions to d and e masses from a component of $h$ lying in a ${\bf 45}$ are required to avoid exact $SU(5)$ mass relations, there is no expectation of seeing precise Georgi-Jarlskog factors of three in the data.

Before 1998, it was generally believed that, in theories of quark-lepton unification, neutrino mixing angles would be small, as with quark mixing angles. However, in 1998 the SuperKamiokande experiment presented convincing data showing atmospheric neutrino oscillations with $\sin^2 \theta_{23} > 0.8$. It was immediately recognized that this could emerge from an $SU(5)$ theory with fermion masses generated by the Froggatt-Nielsen mechanism based on a $U(1)$ flavor symmetry with equal charges for $\bar{F}_2$ and $\bar{F}_3$ \cite{Elwood:1998kf,Altarelli:1998ns}.  From the perspective of MAFS, however, large leptonic mixing does not result from special charge assignments, but is a direct consequence of $SU(5)$ and the observed hierarchies of quark and charged lepton masses.

\subsection{$SU(5)$ in 5d}
\label{subsec:5d}
Several advantages emerge if the $SU(5)$ gauge bosons propagate in 5d, with one dimension being the $S_1/(Z_2 \times Z_2')$ orbifold \cite {Kawamura:2000ev, Altarelli:2001qj, Hall:2001pg}. Orbifold boundary conditions break the gauge group to $SU(3) \times SU(2) \times U(1)$, and the SM Higgs doublet arises as a zero-mode while no such zero-mode exists for its color triplet partner. In supersymmetric theories, proton decay is absent at dimension 4 and 5, and precision gauge coupling unification can be maintained, even though there is a brane where $SU(5)$ is broken. For questions of flavor, the focus of this paper, new opportunities arise because some quark and lepton superfields can reside in the bulk of the orbifold, while others are restricted to one of the branes \cite{Hall:2001pg, Hebecker:2001wq, Hall:2002ci}. The key point is that if either $T_i$ or $\bar{F}_i$ propagate in the bulk then the corresponding Yukawa couplings $y_{e_i}$ and $y_{d_i}$ are not unified. This results because not all of the components of $T_i$ or $\bar{F}_i$ contain zero modes; rather the zero modes of a generation require a doubling of the bulk fields 
\begin{equation}
T_i \;\rightarrow \; T^+_i(\bar{u}, \bar{e}), \, T^-_i(q) \hspace{0.5in}
\bar{F}_i \;\rightarrow \; \bar{F}^+_i(\bar{d}), \, \bar{F}^-_i(\ell)
\label{eq:TFin5d}
\end{equation}
where the $\pm$ superscripts give the $Z'$ orbifold quantum number of the $SU(5)$ fields and the zero modes are shown in parentheses. Hence, for $SU(5)$ unification, another virtue of the $S_1/(Z_2 \times Z_2')$ orbifold arises from flavor: the Higgs can lie purely in a $\bf{5}$, there is no need to add scalars in a $\bf{45}$.

What locations should we choose for $T_i$ and $\bar{F}_i$?\footnote{When the bulk is chosen, it is understood that we mean these fields to contain both $Z' = \pm$ modes.} Recall that supersymmetry does not allow bulk Yukawa couplings, they must all be located on the branes. Furthermore, if the volume of the bulk is large, as required for precision gauge coupling unification, then the Yukawa coupling involving any bulk field will be volume suppressed by a factor $\epsilon_V \sim 0.1 - 0.03$ \cite{Hall:2002ci}. A glance at the quark and lepton spectra of Fig.  \ref{fig:PanelofYukawas} leads us to diverge from the choices made in the original papers.  Apparently, the top quark is special: its Yukawa coupling is 50 times larger than the next largest, while the 8 remaining Yukawa couplings all very loosely cluster within a spread of about $50^2$. Hence, we consider the possibility that $T_3$ alone resides on the $SU(5)$ brane, while all other matter fields reside in the bulk. Neglecting suppression factors from approximate flavor symmetries, the SM Yukawa couplings take the form
\begin{equation}
Y^U \sim \epsilon_V\begin{bmatrix}
     \epsilon_V^2 & \epsilon_V^2   & \epsilon_V \\
     \epsilon_V^2 & \epsilon_V^2 & \epsilon_V  \\
      \epsilon_V& \epsilon_V& 1 \end{bmatrix}
      \hspace{0.5in}
      Y^D \sim \epsilon_V\begin{bmatrix}
     \epsilon_V^2 & \epsilon_V^2   & \epsilon_V^2 \\
     \epsilon_V^2 & \epsilon_V^2 & \epsilon_V^2  \\
      \epsilon_V& \epsilon_V& \epsilon_V \end{bmatrix}
      \hspace{0.5in}
      Y^E \sim \epsilon_V\begin{bmatrix}
     \epsilon_V^2 & \epsilon_V^2   & \epsilon_V \\
     \epsilon_V^2 & \epsilon_V^2 & \epsilon_V  \\
      \epsilon_V^2& \epsilon_V^2& \epsilon_V \end{bmatrix}
\label{eq:yvol}
\end{equation}
where order unity couplings are omitted from all entries, and these differ between $Y^D$ and $Y^E$. The overall factor of $\epsilon_V$ arises because the Higgs fields reside in the bulk, so that an order unity top quark Yukawa coupling results from a 5d coupling that, like the gauge coupling, approaches strong coupling. Thus, from these volume factors alone, for the heavy two generations we expect $y_b/y_t \sim y_s/ y_b \sim y_\mu/y_\tau \sim V_{cb} \sim \epsilon_V$. Taking $\epsilon_V = 0.03$ gives all four of these observables within a factor of 2 of their measured values. We also expect $y_c/y_t \sim  \epsilon_V^2$, which is off by a factor of 3. It is only for the first generation masses, and mixings to the first generation, that further suppression from approximate flavor symmetries is needed.

Since bulk matter multiplets come in pairs, with $Z' = \pm 1$, this configuration of matter requires 11 superfields, and hence there are 11 spurions associated with the breaking of the Maximal Abelian Flavor Symmetry. Here, however, we assume that each pair of bulk fields transforms the same way under the flavor group, so that there are just 6 independent spurions and the entries $Y^{U, D, E}_{ij}$ of (\ref{eq:yvol}) are to be multiplied by $\epsilon^T_{5i} \epsilon^T_{5j}, \epsilon^T_{5i} \epsilon^{\bar{F}}_{5j}, \epsilon^{\bar{F}}_{5i} \epsilon^T_{5j}$, where $\epsilon_5$ denotes MAFS spurions in the 5d theory. The heaviest two generations are consistent with $(\epsilon_5)^{T, \bar{F}}_{2,3} \sim 1$ and the first generation is successfully accounted for by $(\epsilon_5)^{T, \bar{F}}_1 \sim 0.07$.

\section{$SO(10)$ Unification}
\label{sec:10}

The unique unified gauge group of minimal rank where all the observed quarks and leptons of a single generation lie in a single irreducible multiplet is $SO(10)$ \cite{Fritzsch:1974nn, Georgi:1974my}; each generation fills out a $\bf{16}$, the smallest spinor representation, $\psi_i(q_i, \bar{u}_i, \bar{d}_i, \ell_i, \bar{e}_i, \bar{\nu}_i)$.
With this set of fermions, the most general Yukawa interactions that generate the charged fermion masses are 
\begin{equation}
    \mathcal{L}_{SO(10), \psi} \; = \; {\cal Y}_{ij}^\alpha \; \psi_i \psi_j \, H^\alpha \; + \; \rm{h.c.}. 
     \label{eq:L10psi}
\end{equation}
The SM Higgs doublet lies in a linear combination of fields, $H^{10,126}$, in tensor representations $({\bf10}, {\bf126})$. This is much simpler than (\ref{eq:L5Y}) for $SU(5)$; the MAFS is reduced to $U(1)^3$, offering the prospect that flavor hierarchies can be described by just three symmetry breaking parameters, $\epsilon^\psi_i$. However, this leads to $Y^{U,D,E}_{ij} = C^{U,D,E}_{ij} \,\epsilon^\psi_i \epsilon^\psi_j$ which, for the third generation, leads to the expectation that $y_t \approx y_b \approx y_\tau$. We cannot expect that the hierarchy between $t$ and $b / \tau$, which is nearly two orders of magnitude, can be accounted for by differences between the order unity coefficients $C^U_{33}$ and $C^{D,E}_{33}$.  One possibility is that this hierarchy emerges from the embedding of SM Higgs in the doublets of $H^\alpha$; however, here we insist that hierarchies arise from MAFS that act only on fermions.

A second very clear problem for MAFS in $SO(10)$ lies in the comparison of the up-type quark and neutrino masses, which should both have mass ratios between generations governed by $(\epsilon^\psi_3)^2 : (\epsilon^\psi_2)^2 : (\epsilon^\psi_1)^2$.  For $t:c:u$ this is $1: 3.3 \times 10^{-3} : 6.6 \times 10^{-6}$, while barely any hierarchy is discernible in the neutrino masses and mixings; a significant step back from the success of MAFS in $SU(5)$.

In applying MAFS to $SO(10)$ we must abandon models with minimal fermions, $\psi_i$. For most of this section we take the SM Higgs to lie in spinor representations, $\alpha$, rather than in tensor representations of $SO(10)$. Since $\psi_i \psi_j \, H^\alpha$ interactions are then forbidden, $\psi_i$ must couple to $H^\alpha$ via some new heavy fermions $X^A_i$ that are in tensor representations, $A$, giving a flavor sector
\begin{equation}
    \mathcal{L}_{SO(10), \psi X} \; = \; {\cal Y}_{ij}^{A \alpha} \; \psi_i X^A_j \, H^\alpha \; + \; M_{X^A_i} X^A_i X^A_i \; + \; \rm{h.c.}, 
     \label{eq:L10psiX}
\end{equation}
where we have chosen a mass basis for $X^A_i$. Theories of this form were introduced to relate predictions for the top quark and Higgs boson masses to precise gauge coupling unification, while also solving the strong CP problem via parity \cite{Hall:2018let, Hall:2019qwx}. Furthermore, a detailed analysis of flavor and leptogenesis was performed \cite{Carrasco-Martinez:2025zus}; however, no flavor symmetry was introduced and higher-dimensional operators played an important role. Here, we explore the consequences of MAFS, which makes all higher-dimensional operators subdominant.

Integrating out the heavy $X^A_i$ states then generates dimension-5 operators of the form $ \psi_i \psi_j H^\alpha H^\beta$ via the seesaw diagram shown in Fig. \ref{fig:DirSS}. We assume that the scalar multiplets $H^\alpha$ contain vevs that break $SU(2)_L$, $v_L$, as well as vevs that are SM singlets, $V \gg v_L$. These operators then generate seesaw masses for quarks and charged leptons, neutrinos and right-handed neutrinos. Dropping all indices, these mass terms take the form
\begin{equation}
    m^{U,D,E}\; \sim \; {\cal Y}^2 \; \frac{ V v_L}{M_X},  \hspace{0.5in}  m^\nu\; \sim \; {\cal Y}^2 \; \frac{v_L^2}{M_X},
   \hspace{0.5in}  m^{\bar \nu} \; \sim \; {\cal Y}^2 \; \frac{V^2}{M_X}. 
     \label{eq:10seesaw}
\end{equation}
In the cases of $G = SU(3) \times SU(2) \times U(1)$ and $G=SU(5)$ the masses $m^{U,D,E}$, $m^\nu$ and $m^{\bar \nu}$ arise from different operators. In $SO(10)$, with the SM Higgs in spinor representations, all these masses arise from a unified mechanism. The large hierarchy of $m_{\bar{\nu}} :m^{U,D,E}: m^\nu \sim V^2 : V v_L : v_L^2$ arises from the use of the different vevs in $H^\alpha$. 

The introduction of the $X_i$ states increases the MAFS, with $\epsilon^\psi_i \rightarrow \epsilon^\psi_i, \epsilon^X_i$. However, $\epsilon^X_i$ do not appear in any of the seesaw masses of (\ref{eq:10seesaw}), as they cancel between ${\cal Y}^2$ and $M_X$. Thus, when $t,b,\tau$ masses all arise from this seesaw, we are still unable to understand the hierarchy between $t$ and $b, \tau$. However, the flavor sector of (\ref{eq:L10psiX}) allows another option for quark and charged lepton masses. For example, if $X^A$ includes a {\bf 10} then the $SU(5)$ decomposition is $\psi_i(10 + \bar{5} + 1) + X_i(5 + \bar{5})$ and a large mass term $\psi_3(\bar{5}) X_3(5)$ leads to a light third generation $\psi_3(10) + X_3(\bar{5})$. The Yukawa coupling for $b/\tau$ arises from the first term in (\ref{eq:L10psiX}), as shown by the diagram on the left of Fig. \ref{fig:DirSS}, while the $t/\nu_\tau$ masses arise from the seesaw term of (\ref{eq:10seesaw}).

More generally, consider some  component $f$ of $\psi$; $f = q, \bar{u}, \bar{d}, \ell$ or $\bar{e}$. As in (\ref{eq:10seesaw}), for simplicity we suppress generation indices. The multiplets $X$ are taken to include components $F$ and $\bar{F}$, where $f$ and $F$ have the same SM gauge quantum numbers. These states receive mass terms from both $M_X$ and ${\cal Y} V$ in (\ref{eq:L10psiX}), of the form
\begin{equation}
    (M_X \, F + {\cal Y} V\, f ) \, \bar{F}. 
     \label{eq:Ffcombo}
\end{equation}
One combination of $F$ and $f$ becomes heavy while the orthogonal combination is the observed SM state. If $M_X \geq {\cal Y} V$ the heavy state is mainly $F$ and integrating it out generates the first term in (\ref{eq:10seesaw}) via the seesaw mechanism. However, if $M_X \leq {\cal Y} V$, the heavy state is mainly $f$ and the SM state is mainly $F$ and hence lies dominantly in $X$. In this case there is no seesaw, the SM Yukawa coupling arises directly from the Yukawa coupling ${\cal Y}^A_{ij}$ of (\ref{eq:L10psiX}). Using the MAFS, ${\cal Y}^A_{ij} \approx \epsilon^\psi_i \epsilon^X_j$, so that these direct contributions to SM Yukawa couplings depend on $\epsilon^X_i$ as well as $\epsilon^\psi_i$. If $q_3$ and $\bar{u}_3$ lie in $\psi_3$, the top quark mass arises from the seesaw, giving $y_t \approx \epsilon^\psi_3 \epsilon^\psi_3$, so that $\epsilon^\psi_3 \sim 1$. However, if $\bar{b}$ and $\tau$ lie dominantly in $X_3$, there is a direct contribution to the $D$ and $E$ Yukawa coupling matrices
\begin{equation}
    (Y^D_{dir}, Y_{dir}^{E \;^T})_{ij}  \; \approx \; \epsilon^\psi_i \, \delta_{j3}\epsilon^X_3. 
     \label{eq:Ydirect}
\end{equation}
Since $\epsilon^\psi_{1,2} \ll \epsilon^\psi_3$, the bottom quark and tau lepton Yukawa couplings are
\begin{equation}
    y_{b, \tau} \; \approx \; \epsilon^\psi_3 \epsilon^X_3 \hspace{1in} \epsilon^\psi_3 \sim 1, \;\epsilon^X_3 \sim 10^{-2}. 
     \label{eq:btaudirect}
\end{equation}
This is the key that allows MAFS to be applied to $SO(10)$. It could be that various SM multiplets of the first two generations lie dominantly in $X_{1,2}$, so that SM Yukawa couplings also involve $\epsilon^X_{1,2}$. However, below we show that this is not necessary and all flavor parameters can be described in terms of $\epsilon^\psi_{1,2,3}$ and $\epsilon^X_3$.

Extending (\ref{eq:Ffcombo}) to three generations, the mass terms for lepton doublets in the MAFS approximation take the form
\begin{equation}
    (\epsilon^X_3 \epsilon^X_3 M_X \, L_3 + \epsilon^X_3 \epsilon^\psi_i V \, \ell_i) \, \bar{L}_3. 
     \label{eq:Ffcomboell}
\end{equation}
Taking the direct limit, $\epsilon^\psi_3 \, V > \epsilon^X_3 M_X$, the heavy mass eigenstate that marries $\bar{L}_3$ is
\begin{equation}
    L'_3 \approx \ell_3 + \left(\frac{M_X}{V} \frac{\epsilon^X_3}{\epsilon^\psi_3}\right) L_3 +  \frac{\epsilon^\psi_2}{\epsilon^\psi_3} \ell_2 + \frac{\epsilon^\psi_1}{\epsilon^\psi_3} \ell_1. 
     \label{eq:L'3}
\end{equation}
The SM lepton doublets, $\ell'_i$, are the three combinations of $(L_3, \ell_i)$ orthogonal to $L'_3$. Later, it will be very important to recall that the projection from $\ell_i$ to $\ell'_j$, and similarly from $\bar{d}_i$ to $\bar{d}'_j$, is given to leading order by
\begin{equation}
 \ell_i \approx \Omega_{ij} \, \ell'_j, \hspace{0.5in} \bar{d}_i \approx \Omega_{ij} \, \bar{d}'_j;  \hspace{0.5in}    \Omega \approx \begin{pmatrix}
         1&0&0\\
         0&1&0\\
         \frac{\epsilon^\psi_1}{\epsilon^\psi_3}&\frac{\epsilon^\psi_2}{\epsilon^\psi_3}&\frac{\epsilon^X_3}{\epsilon^\psi_3}\frac{M_X}{V}\\
     \end{pmatrix},
     \label{eq:Omega}
\end{equation}
where $M_{X_i} \approx \epsilon^X_i\epsilon^X_i M_X$ and X refers to the $SO(10)$ multiplet containing $L, \bar{D}$. 

Two stages of spontaneous symmetry breaking occur to break $SO(10)$ to the SM gauge group
\begin{equation}
 SO(10) \;\;\stackrel{V_{10}}{\longrightarrow} \;\; G_I \;\; \stackrel{V}{\longrightarrow} \;\;SU(3) \times SU(2) \times U(1)
   \label{eq:GSB}
\end{equation}
At scale $V_{10}$, $SO(10)$ is broken to $G_I$, the intermediate gauge group. This could be a left-right symmetric group, such as $SU(4) \times SU(2)_L \times SU(2)_R$ or $SU(3) \times SU(2)_L \times SU(2)_R \times U(1)_{B-L}$, or the smaller group $SU(3) \times SU(2)_L \times U(1) \times U(1)$ which can emerge from the breakdown via $SU(5) \times U(1)$. Second, $G_I$ is broken to the SM gauge group at scale $V$. 
It could be that the mass scale $M_X$ is a new independent mass scale of the theory, for example softly breaking some symmetry.  However, we find that fits to the quark and lepton masses lead to values of $M_X$ close to the scale $V$.  Hence, we take $X$ masses to arise spontaneously from $\lambda_{X_i} X_i X_i \,\phi$, with $\langle \phi \rangle \, \sim V$. In this case, with $\lambda_{X_i} \approx \epsilon^X_i \epsilon^X_i$, we obtain $M_X \approx V$, so the factor of $(M_X/V)$ in (\ref{eq:Omega}) can be dropped.

\subsection{A Minimal Model with Spinor SM Higgs}
The class of $SO(10)$ theories we have introduced appears wide since many choices can be made for $X^A_i$ and $H^\alpha$ multiplets. However, since the masses of the $X^A_i$ states are at least a few orders of magnitude below $V_{10}$, perturbative gauge coupling unification restricts the size of these multiplets. We study the minimal set $X^{10}_i, X^{45}_i$, the two smallest tensor representations. We take the SM Higgs to be embedded in the two smallest spinor representations, $H(16), H'(144)$, although we will find that including $H'(144)$ is not necessary. The flavor sector of this minimal theory is
\begin{equation}
    \mathcal{L}_{SO(10), \rm{min}} \; = \; {\cal Y}_{ij}^{A,\alpha} \; \psi_i X_j^A \, H^{\alpha} \; + \; \lambda_i^A \;X_i^A X_i^A \phi\; + \; \rm{h.c.}, \hspace{0.25in} A = 10,45; \;\; \alpha = 16,144. 
     \label{eq:L10Ymin}
\end{equation}

The $SO(10)$ group theory of these multiplets forces a separation between the generation of $Y^{D,E}$ and $Y^{U,\nu}$. 
Since $X^{10}$ decomposes into $5 + \bar{5}$ under $SU(5)$, it contains vector-like fermions $D, \bar{D}, L, \bar{L}$, and hence $X^{10}$ is relevant for generating $Y^{D,E}$. On the other hand, $X^{45}$ decomposes into $ 1 + 10 + \bar{10} + 24$ and hence contains $S, Q, \bar{Q}, U, \bar{U}, ...$ so that it is relevant for $Y^{U,\nu}$. In the $SU(5)$ theory of (\ref{eq:L5Y}), the relation $Y^E = (Y^D)^T$ was avoided because the SM Higgs, $h$, had order unity components in both $H^5$ and $H^{45}$.  In the $SO(10)$ case, order unity differences between $Y^E$ and $(Y^D)^T$ emerge from two sources: $h$ may lie in both $H^{16}$ and $H^{144}$, and the vev of $\phi$ may not preserve $SU(5)$ so that $M_E \neq M_D$.

We take ${\cal Y}_{33}^{10,\alpha} \, V \gg \lambda^{10}_i \, V \sim M^{D,E}_3$, ensuring that $b$ and $\tau$ masses arise directly as described above, giving (\ref{eq:btaudirect}), where it is understood that $\epsilon^X_i$ refer to the $X^{10}_i$ states. Masses for all other quarks and leptons are taken to arise via seesaw diagrams that generate dimension-5 operators.
\begin{figure}
\begin{tikzpicture}[>=stealth,thick,font=\small,
                    baseline=(current bounding box.center)]
\tikzset{
  fermion/.style={thick},
  scalar/.style={dashed,thick},
  cross/.style={cross out, draw, minimum size=4pt, inner sep=0pt}
}

\draw[fermion,->] (-2,0) -- (-1,0);
\draw[fermion,-] (-1,0) -- (0,0);
\draw[fermion,-] (0,0) -- (1,0);
\draw[fermion,<-] (1,0) -- (2,0);

\node[below] at (-2,0) {$\psi_3$};
\node[below] at (2,0) {$X^{10}$};
\node[below] at (0,0) {$\mathcal{Y}^{10,\alpha}$};

\draw[scalar,-] (0,0) -- (0,1);
\draw[scalar,<-] (0,1) -- (0,2);
\node[above] at (0,2) {$H^{\alpha}$};

\begin{scope}[shift={(8,0)}]
\draw[fermion,->] (-3,0) -- (-2.25,0);
\draw[fermion,-] (-2.25,0) -- (-1.5,0);
\draw[fermion,-] (-1.5,0) -- (-0.75,0);
\draw[fermion,<-] (-0.75,0) -- (0,0);

\draw[fermion,->] (0,0) -- (0.75,0);
\draw[fermion,-] (0.75,0) -- (1.5,0);
\draw[fermion,-] (1.5,0) -- (2.25,0);
\draw[fermion,<-] (2.25,0) -- (3,0);
\node[below] at (-3,0) {$\psi_i$};
\node[below] at ( 3,0) {$\psi_j$};
\node[below] at (0,0) {$\lambda_A$};
\node[below] at (-1.5,0) {$\mathcal{Y}^{A\alpha}$};
\node[below] at (1.5,0) {$\mathcal{Y}^{A\beta}$};
\draw[scalar,-] (-1.5,0) -- (-1.5,1);
\draw[scalar,<-] (-1.5,1) -- (-1.5,2);
\node[above] at (-1.5,2) {$H^{\alpha}$};
\draw[scalar] (0,0) -- (0,2);
\node[cross] at (0,2) {};
\node[above] at (0,2) {$\braket{\phi}$};
\draw[scalar,-] (1.5,0) -- (1.5,1);
\draw[scalar,<-] (1.5,1) -- (1.5,2);
\node[above] at (1.5,2)
  {$H^{\beta}$};

\node[above] at (-0.75,0)
  {$X^{A}$};

\node[above] at (0.75,0)
  {$X^{A}$};

\end{scope}
\end{tikzpicture}
\caption{$b$ and $\tau$ masses arise directly from the Yukawa coupling on the left. All other quark and lepton masses arise from the seesaw diagram on the right; when external fermions are $\bar{d}$ or $\ell$, they must be projected onto the light SM states using (\ref{eq:Omega}). In the minimal model $A = 10, 45$ and $\alpha = 16, 144$.}
\label{fig:DirSS}
\end{figure}
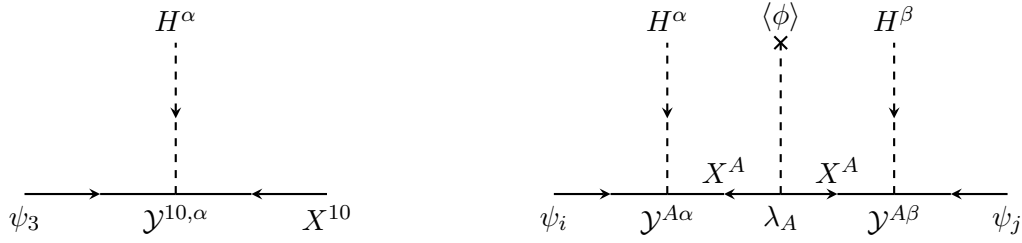
Seesaw contributions to $Y^{D,E}$ arise from the diagram of Figure \ref{fig:DirSS} when one of the $H^{16,144}$ legs is set to the vev $V$ and the other becomes the SM Higgs
\begin{equation}
    (Y^D_{\text{ss}})_{i\ell} = \left( \alpha \, {\cal Y}^{16}_{ij}  + \beta \;{\cal Y}^{144}_{ij} + \gamma \,C_D  \; {\cal Y}^{144}_{ij} \right) \frac{1}{\lambda_j \langle \phi \rangle C'_D} \left({\cal Y}^{T \,16}_{jk} V^{16} + C''_D {\cal Y}^{T \,144}_{jk} V^{144}\right)\, \Omega_{k\ell}.
     \label{eq:YDseesaw}
\end{equation}
\begin{equation}
    (Y^E_{\text{ss}})_{i\ell} \; = \Omega^T_{\ell k} \left( \alpha \, {\cal Y}^{16}_{kj}  + \beta \;{\cal Y}^{144}_{kj} + \gamma \,C_E  \; {\cal Y}^{144}_{kj} \right) \frac{1}{\lambda_j \langle \phi \rangle C'_E} \left({\cal Y}^{T \,16}_{ji} V^{16} + C''_E {\cal Y}^{T \,144}_{ji} V^{144}\right).
     \label{eq:Yeseesaw}
\end{equation}
The SM Higgs is embedded via $H^{16} = V^{16} + \alpha \, h \,+ ...$ and $H^{144} = V^{144} + \beta \, h \,+ \gamma h \, +...$, where $(\beta \, h \, , \gamma h)$ lie in the $SU(5)$ $(5, 45)$ pieces of $H^{144}$, and $|\alpha|^2 + |\beta|^2 + |\gamma|^2 = 1$. It is understood that ${\cal Y}^{16,144}$ and $\lambda$ refer to the Yukawa couplings of $X^{10}$ and $k=1,2$. The Clebsch factors $C, C', C''$ differ between the $D$ and $E$ sectors, and depend on the intermediate gauge group $G_I$.  
Provided $C'_D \neq C'_E$ the $H^{144}$ multiplet is not necessary; we include it to stress that our results apply to a wide range of models. The matrix $\Omega$ results from applying the projections of (\ref{eq:Omega}). 
Only two eigenvalues arise from the seesaw. The total SM Yukawa matrices are $Y^{D,E} = Y^{D,E}_{\text{dir}} + Y^{D,E}_{\text{ss}}$. In the MAFS approximation, where all couplings ${\cal Y}_{ij}, \lambda_i$ are given by their dependence on the $\epsilon$ parameters and the order unity coefficients $\alpha, \beta, \gamma, C, C', C''$ are dropped, these become
\begin{equation}
Y^D \approx \begin{pmatrix}
     \epsilon_1^2&\epsilon_1\epsilon_2&\epsilon_1 \epsilon_X\\
     \epsilon_1\epsilon_2&\epsilon_2^2&\epsilon_2\epsilon_X\\
     \epsilon_1\epsilon_3&\epsilon_2\epsilon_3&\epsilon_3\epsilon_X
    \end{pmatrix}, \quad \quad
    Y^E \approx \begin{pmatrix}
     \epsilon_1^2&\epsilon_1\epsilon_2&\epsilon_1\epsilon_3\\
     \epsilon_1\epsilon_2&\epsilon_2^2&\epsilon_2\epsilon_3\\
     \epsilon_1\epsilon_X&\epsilon_2\epsilon_X&\epsilon_3\epsilon_X
    \end{pmatrix}
    \label{eq:YDEMAFS}
\end{equation}
where, for clarity of notation, we drop the $\psi$ superscript from $\epsilon_i^\psi$ and $\epsilon_X$ is the MAFS parameter for $X^{10}_3$. The entries proportional to $\epsilon_X$ receive comparable contributions from both direct and seesaw terms.

The Yukawa couplings in the up sector, $Y^U$, and the neutrino sector, $Y^\nu$ relevant for leptogenesis, arise entirely from the seesaw diagram in Figure \ref{fig:DirSS} when $X^{45}$ is integrated out, one of the $H^{16,144}$ legs is set to the vev $V$ and the other becomes the SM Higgs 
\begin{equation}
    Y^U_{ik} \; = \; \left( \alpha \, {\cal Y}^{16}_{ij}  + \beta \;{\cal Y}^{144}_{ij} + \gamma \,C_U  \; {\cal Y}^{144}_{ij} \right) \frac{1}{\lambda_j \langle \phi \rangle C'_U} \left({\cal Y}^{T \,16}_{jk} V^{16} + C''_U {\cal Y}^{T \,144}_{jk} V^{144}\right).
     \label{eq:YUseesaw}
\end{equation}
\begin{equation}
    Y^\nu_{i \ell} \; = \Omega^T_{ij} \left( \alpha \, {\cal Y}^{16}_{jk}  + \beta \;{\cal Y}^{144}_{jk} + \gamma \,C_\nu  \; {\cal Y}^{144}_{jk} \right) \frac{1}{\lambda_k \langle \phi \rangle C'_\nu} \left({\cal Y}^{T \,16}_{k \ell} V^{16} + C''_\nu {\cal Y}^{T \,144}_{k \ell} V^{144}\right).
     \label{eq:Ynuseesaw}
\end{equation}
where ${\cal Y}^{16,144}$ and $\lambda$ refer to the Yukawa couplings of $X^{45}$ and $k=1,2,3$. Note that $\Omega$ appears in $Y^\nu$ but not in $Y^U$. In the MAFS approximation, these become
\begin{equation}
Y^U \approx \begin{pmatrix}
     \epsilon_1^2&\epsilon_1\epsilon_2&\epsilon_1 \epsilon_3\\
     \epsilon_1\epsilon_2&\epsilon_2^2&\epsilon_2\epsilon_3\\
     \epsilon_1\epsilon_3&\epsilon_2\epsilon_3&\epsilon_3\epsilon_3
    \end{pmatrix}, \quad \quad
    Y^\nu \approx \begin{pmatrix}
     \epsilon_1^2&\epsilon_1\epsilon_2&\epsilon_1\epsilon_3\\
     \epsilon_1\epsilon_2&\epsilon_2^2&\epsilon_2\epsilon_3\\
     \epsilon_1\epsilon_X&\epsilon_2\epsilon_X&\epsilon_3\epsilon_X
    \end{pmatrix}.
    \label{eq:YUnuMAFS}
\end{equation}
The $\bar{\nu} \; (\nu)$ mass matrices occur from the seesaw diagrams of Figure \ref{fig:DirSS} when both $H^{16,144}$ legs are set to the vev $V \; (v_L)$ giving, in the MAFS approximation
\begin{equation}
m_{\bar{\nu}} \approx \begin{pmatrix}
     \epsilon_1^2&\epsilon_1\epsilon_2&\epsilon_1 \epsilon_3\\
     \epsilon_1\epsilon_2&\epsilon_2^2&\epsilon_2\epsilon_3\\
     \epsilon_1\epsilon_3&\epsilon_2\epsilon_3&\epsilon_3\epsilon_3
    \end{pmatrix} V, \quad \quad
    m_\nu \approx \begin{pmatrix}
     \epsilon_1^2&\epsilon_1\epsilon_2&\epsilon_1\epsilon_X\\
     \epsilon_1\epsilon_2&\epsilon_2^2&\epsilon_2\epsilon_X\\
     \epsilon_1\epsilon_X&\epsilon_2\epsilon_X&\epsilon_X^2
    \end{pmatrix} \frac{v_L^2}{V}.
    \label{eq:mnuMAFS}
\end{equation}

From $Y^{U,D,E}$ of (\ref{eq:YDEMAFS}, \ref{eq:YUnuMAFS}), the quark and lepton masses and the CKM matrix are predicted in terms of the four parameters $(\epsilon_i, \epsilon_X)$. Values of $(\epsilon_i, \epsilon_X)$ that yield each of these 12 observables in the MAFS approximation are shown in Table \ref{tab:SO10eps}. The SM parameters are renormalized at $10^{12}$ GeV for these fits. Below we find that neutrino masses determine $V$ to be of this order. The precise form of the theory above $V$ is unknown; still, scaling to $10^{16}$ GeV likely changes Yukawa couplings by 10\% or less. Approximate central values for these parameters are
\begin{equation}
    \bar{\epsilon}_1 = 0.003, \;\;\;\;  \bar{\epsilon}_2 = 0.02, \;\;\;\; \bar{\epsilon}_3 = 0.7, \;\;\;\;  \bar{\epsilon}_X = 0.01. 
     \label{eq:centraleps}
\end{equation}
The spread in the extracted values of each $\epsilon$ parameter gives a measure of the order unity coefficients required for an understanding of the observables within MAFS. The deviations of $\epsilon_{i,X}$ from the central values of (\ref{eq:centraleps}) are always less than a factor of 2, and often much less, so that the MAFS approximation is remarkably successful. Since $\epsilon_3 \sim 1$, the hierarchical values of these 12 observables can be broadly understood in $SO(10)$ in terms of only 3 $\epsilon$ parameters.

\begin{table}[]
\centering
\small
\setlength{\tabcolsep}{4pt}
\begin{tabular}{|c|c c c|c c c|c c c|c c c|}
\hline
 & $m_t$ & $m_b$ & $m_\tau$ & $m_c$ & $m_s$ & $m_\mu$ & $V_{cb}$ & $V_{ub,td}$ & $V_{us}$ & $m_u$ & $m_d$ & $m_e$ \\
\hline
$\epsilon_3$ & 0.7 & 0.7 & 0.7 &  &  &  & 0.7 & 0.7 &  &  &  &  \\
\hline
$\epsilon_2$ &  &  &  & 0.040 & 0.012 & 0.024 & 0.032 &  & 0.020 &  &  &  \\
\hline
$\epsilon_1$ &  &  &  &  &  &  &  & 0.0048 & 0.0045 & 0.0018 & 0.0027 & 0.0017 \\
\hline
$\epsilon_X$ &  & 0.010 & 0.013 &  &  &  &  &  &  &  &  &  \\
\hline
\end{tabular}
\caption{Values of $(\epsilon_i, \epsilon_X)$ in the minimal $SO(10)$ model, extracted from quark and charged lepton masses and the CKM matrix in the MAFS approximation.}
\label{tab:SO10eps}
\end{table}

In $SO(10)$ theories that have the minimal set of fermions, $\psi$,  $Y^{U,D,E}, m_\nu$ the flavor matrices are all proportional to $\epsilon_i \epsilon_j$. However, by adding $X^{10,45}$ fermions only the most hierarchical, $Y_U$, takes this form.  In several entries of $Y^D, Y^E$ and $m_\nu$
the order-unity $\epsilon_3$ is replaced by $\epsilon_X \sim 0.01$, as seen in (\ref{eq:YDEMAFS}), (\ref{eq:YUnuMAFS}) and (\ref{eq:mnuMAFS}).
This not only resolves the conventional $SO(10)$ puzzle of $y_{b,\tau}/y_t$, it explains the absence of any clear hierarchies in the neutrino mass matrix. Because $\epsilon_X$ is intermediate in size between $\epsilon_1$ and $\epsilon_2$, with $\epsilon_2: \epsilon_X : \epsilon_1 \sim 2 : 1: 1/3$, we see from (\ref{eq:mnuMAFS}) that the neutrino masses and mixings are predicted to show anarchy rather than hierarchy. Given the unknown order unity coefficients, no precise predictions are possible; however, the spectrum is normal ordered and there is no particular suppression of $m_{ee}$ relevant for $0\nu$ double beta decay.
 
Given the uncertainties from the order unity factors, $C$, that have been dropped, we cannot be sure which of $\epsilon_2$ and $\epsilon_X$ is larger. It is convenient to assemble $\epsilon_1, \epsilon_2, \epsilon_X$ into $\tilde{\epsilon}_i$, arranged such that $\tilde{\epsilon}_1 = \epsilon_1$ and $\tilde{\epsilon}_2$ ($\tilde{\epsilon}_3$) is the smaller (larger) of $\epsilon_2$ and $\epsilon_X$. The $\nu$ and $\bar{\nu}$ mass matrices, as well as the neutrino Yukawa matrix, can then be written in the simple form 
\begin{equation}
m_{\bar{\nu}_{ij}} \approx \epsilon_i \epsilon_j \, V, \quad \quad \quad
    m_{\nu_{ij}} \approx \tilde{\epsilon}_i \tilde{\epsilon}_j \frac{v_L^2}{V}, \quad \quad \quad Y^\nu_{ij} \approx \tilde{\epsilon}_i \epsilon_j.
    \label{eq:mnuYnuMAFS}
\end{equation}
The scale of neutrino masses allows us to predict the intermediate scale of gauge symmetry breaking to be far below the scale of $SO(10)$ breaking
\begin{equation}
    V \approx \frac{\tilde{\epsilon}_3^2 \;v_L^2}{m_{\nu_3}} \simeq 5 \times 10^{11} \text{GeV} \; \left(\frac{\tilde{\epsilon}_3}{0.03}  \right)^2 \left(\frac{0.05 \,\text{eV}}{m_{\nu_3}}\right).
     \label{eq:Vfrom mnu}
\end{equation}
Gauge coupling unification then prefers an intermediate gauge group of $G_I = SU(3) \times SU(2)_L \times SU(2)_R \times U(1)_{B-L}$ and, furthermore, parity may be imposed to solve the strong CP problem \cite{Hall:2018let, Hall:2019qwx, Carrasco-Martinez:2025zus}.

\begin{figure}
    \centering
    \includegraphics[width=1.0\linewidth]{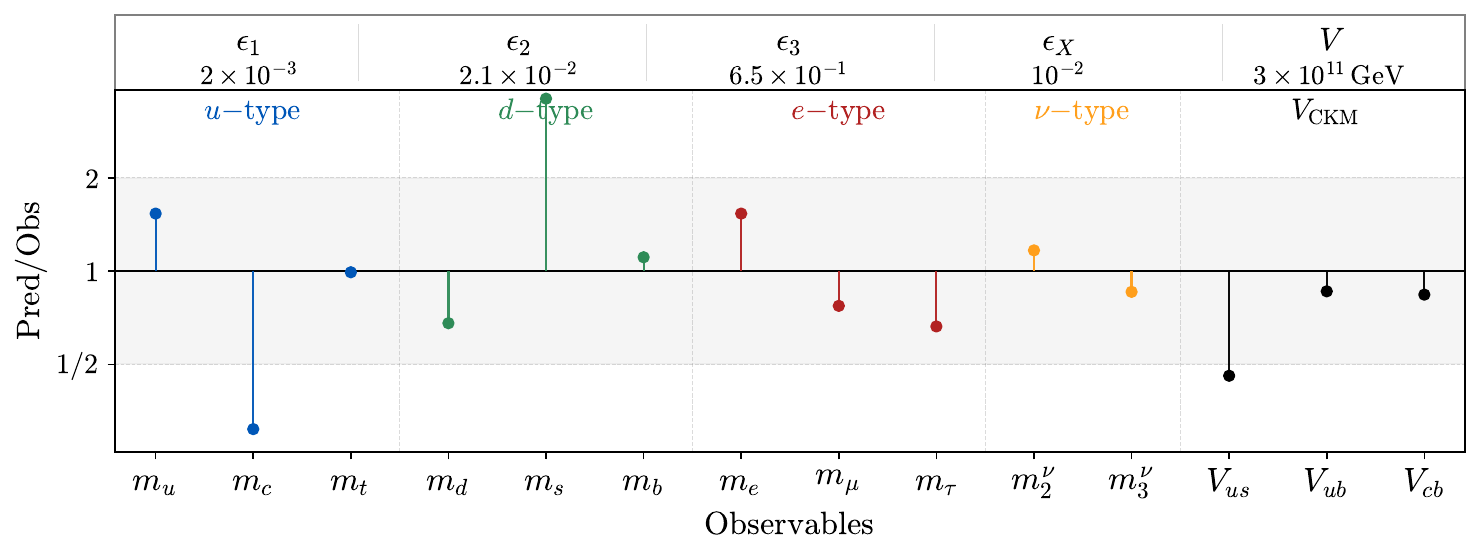}
    \caption{Results of a fit in the $SO(10)$ theory to 14 observables (quark masses, lepton masses, and CKM parameters) with 5 free parameters $\epsilon_i, \epsilon_X, V$, using the MAFS approximation.}
    \label{fig:fitnoYB}
\end{figure}

In Fig.  \ref{fig:fitnoYB} we show the result of fitting 14 observables $O_i$, the masses $m_{u,c,t}, m_{d,s,b}, m_{e,\mu,\tau}$,  $m_{\nu_2, \nu_3}$ and CKM matrix elements $,V_{us}, V_{ub}, V_{cb}$, to the 5 parameters $(\epsilon_i, \epsilon_X,V)$ by minimizing $d^2 = \sum_i \ln^2(O_i/O_{i,obs})$. We use the MAFS approximation where each observable is approximated by its dependence on $\epsilon_i, \epsilon_X$.

\subsection{Generality of Results}

The key to obtaining realistic masses in the previous subsection was that, unlike all the other quarks and leptons, $(\bar{d}_3, \ell_3)$ lie predominantly outside the 16-dimensional multiplets $\psi_i$. This generalizes well beyond the minimal model. For example, the SM Higgs may lie in a combination of tensor (10, 126) and spinor (16) multiplets, and the only relevant fermions beyond $\psi_i$ might be a single 10 multiplet, $X$, with Yukawa interactions
\begin{equation}
    \mathcal{L}_{SO(10)} \; = \; {\cal Y}_{ij}^{10} \; \psi_i \psi_j \, H_{10}  + {\cal Y}_{ij}^{126} \; \psi_i \psi_j \, H_{126}^*  + {\cal Y}_i \; \psi_i X \, H_{16}+ \; \lambda \;X X \; \phi \; + \; \rm{h.c.}, 
     \label{eq:L10Y2}
\end{equation}
where $\phi$ is an $SO(10)$-singlet scalar.
In the MAFS approximation
\begin{equation}
{\cal Y}_{ij}^{10, 126} \approx \epsilon_i \epsilon_j, \qquad {\cal Y}_i \approx \epsilon_i \epsilon_X, \qquad
    \lambda \approx \epsilon_X^2.
    \label{eq:nonminMAFS}
\end{equation}
Taking the $SU(5)$ preserving vevs of $H_{16}$ and $\phi$ to be comparable, $(\bar{d}_3, \ell_3)$ lies predominantly in $X$ if $\epsilon_X \ll \epsilon_3 \sim 1$. 

In the MAFS approximation the quark and lepton mass predictions are the same as in the minimal model:
\begin{itemize}
    \item $\psi_i X \, H_{16}$ of (\ref{eq:L10Y2}) gives the ``direct'' mass terms for $b$ and $\tau$, corresponding to the left diagram in Fig. \ref{fig:DirSS}, and the mixing matrix $\Omega$ of (\ref{eq:Omega}).
    \item $\psi_i \psi_j \, (H_{10}, H_{126}^*)$ of (\ref{eq:L10Y2}) leads to the rest of the Yukawa couplings of the $D,E$ sectors, and all of the Yukawas of the $U, \nu$ sectors, replacing the seesaw diagrams of Fig. \ref{fig:DirSS}. The presence of two Yukawa matrices ensures that the CKM matrix is non-trivial.  Furthermore, while $H_{10}$ leads to identical contributions in D and E sectors, those from $H_{126}$ differ by a factor 3, avoiding the predictions $m_s = m_\mu$ and $m_d = m_e$ from $\psi_i \psi_j \, H_{10}$ alone. These first two bullets imply that $Y^{U,D,E,\nu}$ are given by (\ref{eq:YDEMAFS}) and (\ref{eq:YUnuMAFS}). 
    \item The $\bar{\nu}$ mass matrix is generated by the vev $V$ of the $SU(5)$ singlet component of $H_{126}$ and seesaw neutrino masses arise from integrating out $\bar{\nu}$, giving (\ref{eq:mnuMAFS}).  
\end{itemize}

Hence, using MAFS the theory defined by (\ref{eq:L10Y2}) gives the same fit to data, Table \ref{tab:SO10eps} and Fig. \ref{fig:fitnoYB}, as the minimal model, (\ref{eq:L10Ymin}). This results because the SM fermions have the same embedding in $SO(10)$ multiplets, and hence the same $\epsilon_i, \epsilon_X$ factors suppress their couplings. These simple models can be augmented with other fields but, as long as the embedding of SM quarks and leptons remains the same, the flavor predictions will persist. Our MAFS predictions, including the one for leptogenesis below, apply in a wide range of $SO(10)$ models. Of course, once the quark and lepton embedding changes, so will the MAFS predictions.  For example, in the minimal model if all three generations of $(\bar{d}_i, \ell_i)$ lie predominantly in $X^{10}_i$, the predictions will revert to those of the $SU(5)$ theory discussed in section \ref{sec:5}.

\section{Leptogenesis}
\label{sec:lepto}

Neutrino masses may arise from integrating out right-handed neutrinos, $\bar{\nu}_i$, via the seesaw mechanism. 
The Yukawa interactions and mass terms for $\bar{\nu}$ are
\begin{equation}
\label{eq:Lnu}
    \mathcal{L}^\nu_{seesaw} = 
    Y^\nu _{ij} \, \ell_i {\bar \nu}_j h  + \frac{1}{2} M^{\bar{\nu}}_{ij} \, {\bar \nu}_i {\bar \nu}_j.
\end{equation}
The cosmological baryon asymmetry may arise from (\ref{eq:Lnu}) via thermal leptogenesis \cite{Fukugita:1986hr}. As the temperature of the universe drops below the mass of $\bar{\nu}_i$, decays $\bar{\nu}_i \rightarrow \ell h, \ell^\dagger h^\dagger$ generate a baryon asymmetry 
\begin{align}
    \label{eq:YB}
    Y_B = \frac{n_B}{s} = \frac{28}{79} Y_{\rm therm} \; \epsilon_L \, \eta\,   \simeq 1.5 \times 10^{-3} \; \epsilon_L \, \eta\,  ,
\end{align}
where $Y_{\rm therm}$ is the thermal yield of $\bar{\nu}_i$ and $\epsilon_L$ is the lepton asymmetry produced per $\bar{\nu}_i$ decay. The efficiency factor $\eta$ is small when $\bar{\nu}_i$ is near thermal equilibrium at decay. This occurs if the seesaw mass generated by $\bar{\nu}_i$ exchange, $m_i^{SS}$, is larger than about 1 meV, when \cite{Giudice:2003jh}
\begin{align}
    \label{eq:eta}
    \eta \simeq (m_* / m^{SS}_i )^{1.16},
    \hspace{0.75in} m_* \simeq 0.5 \times 10^{-3}  \, \mbox{eV}.
\end{align}
At 1-loop order, when $\bar{\nu}_i$ decays via a virtual heavier $\bar{\nu}_j$ the asymmetry factor $\epsilon_L$ is given by
\begin{align}
    \label{eq:epsL}
    \epsilon_L
    = \frac{1}{8\pi} \frac{\text{ Im}(Y^{\nu^\dagger} Y^\nu)^2_{ji}}{(Y^{\nu^\dagger} Y^\nu)_{ii}} g\left(\frac{M_{\bar{\nu}_i}}{M_{\bar{\nu}_j}}\right).
\end{align}
For $M_{\bar{\nu}_j}$ of order or much greater than $M_{\bar{\nu}_i}$, a good  approximation is $g \simeq (3/2)M_{\bar{\nu}_i}/M_{\bar{\nu}_j}$. 
In MAFS,  the dependence of $Y^\nu _{ij}$ and $M^{\bar{\nu}}_{ij}$ on $\epsilon_a$ depends on the gauge group $G$.  

\subsection{$G = SU(3) \times SU(2) \times U(1)$}

The addition of three right-handed neutrino fields, $\bar{\nu}_i$ leads to the introduction of three more MAFS symmetry breaking parameters, $\epsilon^{\bar{\nu}}_i$, giving 
\begin{equation}
       Y^\nu_{ij} \; = \; C^\nu_{ij} \, \epsilon^\ell_i \epsilon^{\bar \nu}_j,\hspace{0.2in}
       M^{\bar{\nu}}_{ij} \;=\; C^{\bar{\nu}}_{ij} \, \epsilon^{\bar{\nu}}_i \epsilon^{\bar{\nu}}_j \, M.
\label{eq:Ynu}
\end{equation}
The values of $\epsilon^\ell_i$ are given in Table \ref{tab:321eps} with $\epsilon^\ell_3$ ranging from $10^{-2}$ to unity. 
Integrating out the heavy $\bar{\nu}_i$, the seesaw mechanism leads to the dimension 5 operator for light neutrino masses
of (\ref{eq:L321}) or (\ref{eq:L5nu5}) with $C^{(5)}_{ij} = \Sigma_{kl} \; C^\nu_{ik} (1/C^{\bar{\nu}}_{kl}) C^\nu_{lj}$.
As expected, the light neutrino masses scale with powers of $\epsilon^\ell$, as in (\ref{eq:321masses}), and are independent of $\epsilon^{\bar{\nu}}$. 
In the MAFS approximation, the masses of $\bar{\nu}_i$ are $M_{\bar{\nu}_i} \approx (\epsilon^{\bar{\nu}}_i)^2 M$, where $M$ is the scale of lepton number violation, introduced in (\ref{eq:L321}).

\begin{figure}[]
    \centering
    \includegraphics[width=1.0\textwidth]{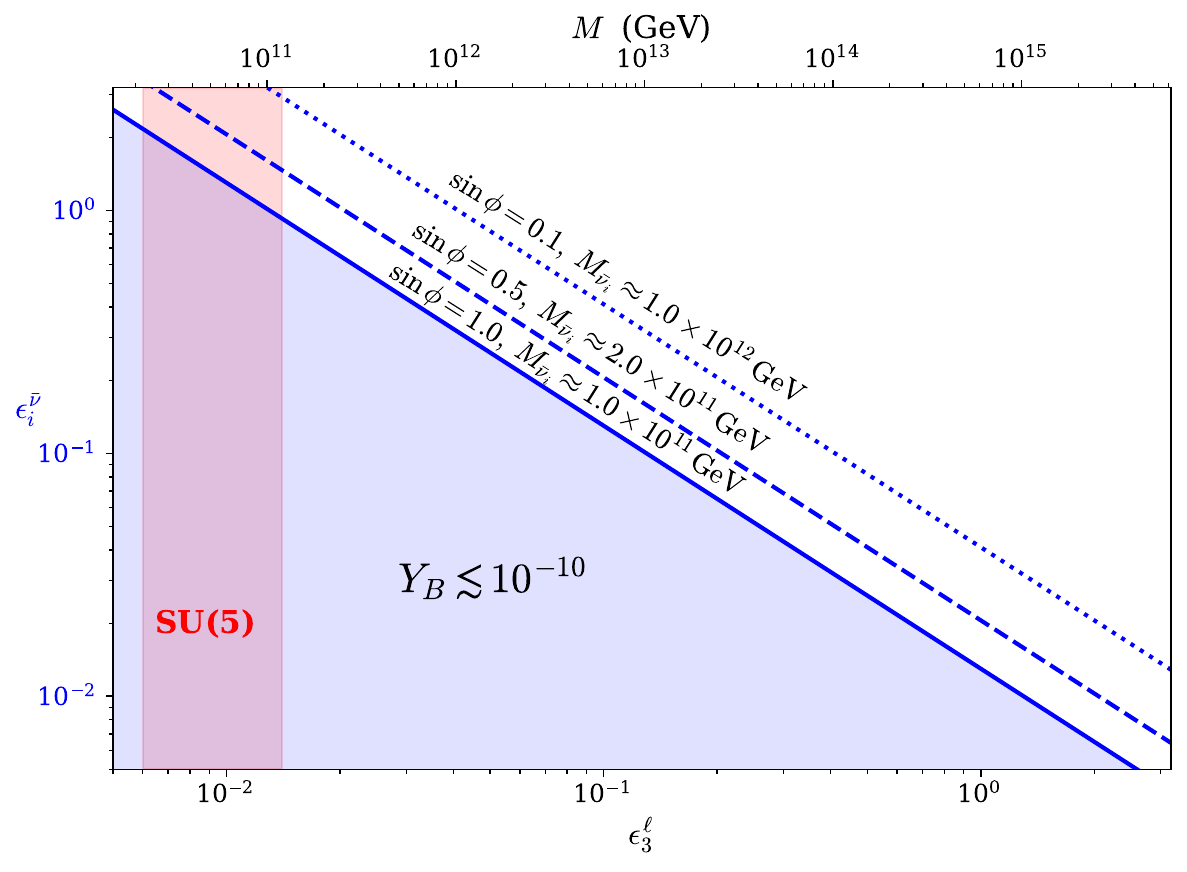} 
    \caption{Thermal leptogenesis from the decay of $\bar{\nu}_i$. Contours of $Y_B = 0.87 \times 10^{-10}$ are shown in blue for $\sin \phi =$ 1 (solid), 0.5 (dashed) and 0.1 (dotted) in the $(\epsilon_3^\ell, \epsilon_i^{\bar{\nu}})$ plane. $Y_B$ is below the observed value in the blue shaded region. For $G = SU(3) \times SU(2) \times U(1)$, the allowed range of $\epsilon_3^\ell$ is $ 0.01 \lesssim \epsilon_3^\ell \lesssim 1$, while for $G=SU(5)$, $\epsilon_3^\ell \approx 10^{-2}$, as shown by the red shaded vertical band. The mass scale $M$ of lepton number violation is related to $\epsilon_3^\ell$ by the observed neutrino mass $m_{\nu_3}$, as shown by the top horizontal axis.}
    \label{fig:YB}
\end{figure}

Since $\epsilon^\ell_3 > \epsilon^\ell_{1,2}$, the MAFS approximation gives a result independent of $\epsilon^{\bar{\nu}}_j$
\begin{align}
    \label{eq:epsLapprox}
    \epsilon_L
    \approx \frac{3}{16\pi} (\epsilon^\ell_3 \epsilon^{\bar \nu}_i)^2 \sin \phi, \qquad i=1 \; \rm{or} \; 2,
\end{align}
where $\phi$ is a combination of phases appearing in $C^\nu_{3j}$. The seesaw mass generated from $\bar{\nu}_i$ exchange is independent of $\epsilon^{\bar{\nu}}_i$ in the MAFS approximation
\begin{align}
    \label{eq:mSS}
    m^{SS}_i
    \approx \frac{(\epsilon^\ell_3 v)^2}{M} \approx m_{\nu_3},
\end{align}
giving a washout factor $\eta \approx 0.5 \times 10^{-2}$. Hence the baryon asymmetry is
\begin{align}
    \label{eq:YB2}
    Y_B  \approx \frac{28}{79} \, Y_{\rm therm}\left( \frac{3}{16\pi} (\epsilon^\ell_3 \epsilon^{\bar \nu}_i)^2 \sin \phi \right) \, \eta \, \approx 0.5 \times 10^{-6} (\epsilon^\ell_3 \epsilon^{\bar \nu}_i)^2 \sin \phi.
\end{align}
Setting $Y_B$ to the observed value of $0.87 \times 10^{-10}$ gives the blue contours in the $(\epsilon^\ell_3, \epsilon^{\bar \nu}_i)$ plane of Fig. \ref{fig:YB}, for various values of $\sin \phi$. The blue shaded region gives $Y_B$ below the observed value and is excluded. The entire region above the solid blue line is allowed for $G = SU(3) \times SU(2) \times U(1)$, but the MAFS framework breaks down for $\epsilon^\ell_3$ or $\epsilon^{\bar \nu}_i$ significantly larger than unity.  For any value of $\epsilon^\ell_3$, the mass scale of lepton number violation, $M$, is determined by (\ref{eq:mSS}), since $m_{\nu_3} \simeq 0.05$ eV, and is shown in the top horizontal axis of Fig. \ref{fig:YB}. Alternatively, using (\ref{eq:mSS}) to determine $\epsilon^\ell_3$ and substituting it into (\ref{eq:YB2}) gives the mass of the decaying right-handed neutrino: $M_{\bar{\nu}_i} \approx  10^{11} \,\mbox {GeV} / \sin \phi$; the blue contours are also lines of fixed $M_{\bar{\nu}_i}$. 

\subsection{$SU(5)$}

For $G = SU(5)$ the analysis is almost the same. The gauge singlet states, $\bar{\nu}_i$, must be added to the theory, together with $\epsilon^{\bar{\nu}}_i$. The only change is that now $\ell_i$ is embedded in $\bar{F}_i$ so  $\epsilon^\ell_i$ are replaced by $\epsilon^{\bar{F}}_i$, given in Table \ref{tab:5epsTF}. In fact, $\epsilon^{\bar{F}}_i$ are the same as $\epsilon^\ell_i$ given in Table \ref{tab:321eps}, except now $\epsilon^\ell_3$ is constrained to be near $10^{-2}$ by the masses for $(t, b, \tau)$. This implies that Fig. \ref{fig:YB} also applies to $SU(5)$, with the restriction to the vertical red band where $\epsilon^\ell_3 \sim10^{-2}$. Most of this band gives $Y_B$ well below $10^{-10}$, so observations require $\epsilon^{\bar \nu}_i, \sin \phi \approx 1$. CP violation in the lepton sector is maximal and both the decaying and exchanged right-handed neutrinos have masses close to $M \approx 10^{11}$ GeV. Hence, assuming maximal CP violation and no flavor symmetry suppression on right-handed neutrinos, {\it the SU(5) theory correctly predicts the order of magnitude of the cosmological baryon asymmetry}.\footnote{We cannot claim any higher accuracy because of the unknown order unity coefficients of (\ref{eq:Ynu}).  In a mass basis for the $\bar{\nu}_i$, $Y_B$ scales as $(C^\nu_{33})^2 C^{\bar{\nu}}_{33}/ C^{\bar{\nu}}_{22}$.  } In terms of SM flavor parameters, the parametric form of this prediction is 
\begin{align}
    \label{eq:YBmax}
    Y_B(SU(5)) \approx\left( \frac{28}{79} Y_{\rm therm} \right) \left( \frac{3 \, y_b^2}{16\pi y_t} \right) \left( \frac{8 \pi v^2}{M_{Pl} \,m_{\nu_3}} \right)^{1.16} \approx   \left( 10^{-3} \right) \left( 10^{-5} \right) \left( 10^{-2} \right) \approx 10^{-10} 
\end{align}
valid for $m_{\nu_3} > m_*$.
For the efficiency factor we have used the approximation $m_* = 8 \pi v^2/ M_{Pl} $, where $M_{Pl}$ is the reduced Planck mass. 

Remarkably, the only flavor parameters appearing in this prediction are the quark and neutrino masses of the heaviest generation. Approximating the exponent appearing in the efficiency factor as 1 rather than 1.16, the factor of $(\epsilon^{\bar{F}_3})^2 \approx y_b^2/y_t$ cancels between numerator and denominator yielding a prediction for $Y_B$ in terms of the fundamental parameters
\begin{align}
    \label{eq:YBmax2}
    Y_B(SU(5)) \approx\left( \frac{28}{79} Y_{\rm therm} \right) \left( \frac{3}{2}\frac{M}{M_{Pl}} \right) \approx   10^{-10}. 
\end{align}
The numerical result follows by taking $M \approx 10^{11}$ GeV from (\ref{eq:5M}), as required by the observed neutrino masses.

While the $SU(5)$ theory has 9 fermion multiplets, only 5 small parameters ($\epsilon^T_{1,2}$ and $\epsilon^{\bar{F}}_i$ ranging from 0.002 to 0.04) are needed to describe all quark and lepton masses and mixings and the cosmological baryon asymmetry. All three right-handed neutrinos have masses of order $10^{11}$ GeV, and thermal leptogenesis requires a reheat temperature after inflation of this scale or larger.

\subsection{$SO(10)$}

In $SO(10)$, with $b, \tau$ lying dominantly in $X_3$ and all other SM states and $\bar{\nu}_i$ lying in $\psi_i$, the right-handed neutrinos, in the MAFS approximation, have Yukawa couplings $Y^\nu_{ij} \approx \tilde{\epsilon}_i \epsilon_j$ and masses
\begin{equation}
    M_{\bar{\nu}_i} \approx \epsilon_i^2 \, V \approx \frac{\epsilon_i^2 \tilde{\epsilon}_3^2 \;v_L^2}{m_{\nu_3}}
    = \left( 5 \times 10^6 \, \rm{GeV},\; 5 \times 10^8 \, \rm{GeV}, \; 2.5 \times 10^{11} \, \rm{GeV}\right)  \; \left(\frac{\tilde{\epsilon}_3}{0.03} \right)^2,
     \label{eq:mnubar}
\end{equation}
for $\epsilon_i = (0.003, 0.03, 0.7)$. We see that $\bar{\nu}_1$ is too light for its decays to generate the baryon asymmetry. On the other hand,  $m_{\bar{\nu}_2}$ is close to the lower bound of $10^9$ GeV required for leptogenesis and we assume that the order unity factors push $m_{\bar{\nu}_2}$ above this bound.
Using (\ref{eq:mnuYnuMAFS}) in the general results for leptogenesis, (\ref{eq:YB}), (\ref{eq:eta}) and (\ref{eq:epsL}), the baryon asymmetry generated by $\bar{\nu}_2$ decay is
\begin{equation}
    Y_B \; \approx \; 4.5 \times 10^{-7} \; \epsilon_2^2 \,\tilde{\epsilon}_3^2 \,\sin 2 \phi \; = \; 4 \times 10^{-13} \; \left(\frac{\epsilon_2}{0.03} \; \frac{\tilde{\epsilon}_3}{0.03} \right)^2 \,\sin 2 \phi,
     \label{eq:YB10}
\end{equation}
where we used the washout factor $\eta \approx (m_*/m_{\nu_3})^{1.16} = 0.5 \times 10^{-2}$.
\begin{figure}
    \centering
    \includegraphics[width=1.0
    \linewidth]{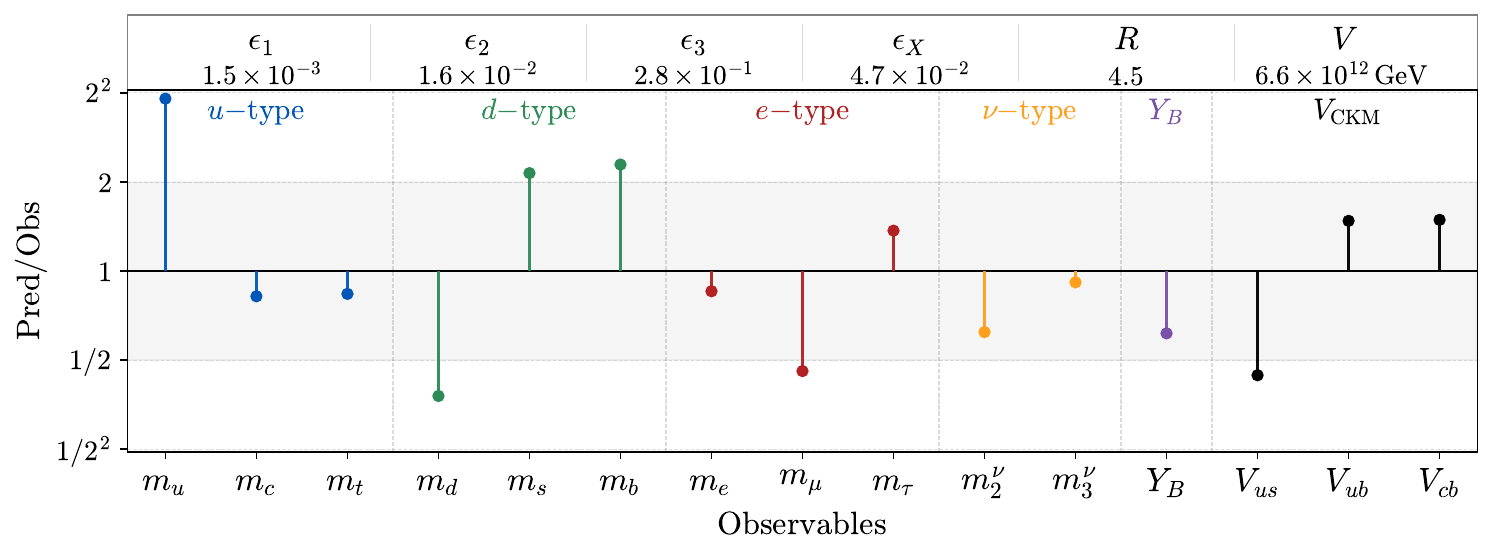}
    \caption{Results of fit to quark masses, CKM parameters, lepton masses and the baryon asymmetry. In addition to the free parameters $\epsilon_i, \epsilon_X, V$ used in the fit of Fig. \ref{fig:fitnoYB}, the ratio of mass scales $R = M^{10}/M^{45}$ is allowed to vary. This significantly improves the fit for $Y_B$ and increases the scale $V$ of intermediate scale gauge symmetry breaking.  }
    \label{fig:fitwithYB}
\end{figure}
For $\epsilon_2 = \tilde{\epsilon}_3 =0.03$ and $\sin 2 \phi=1$, this result for $Y_B$ is too small by a factor of 200. This could simply be due to a pileup of order unity factors or it could be that the minimal model of (\ref{eq:L10Ymin}) is missing some ingredient.  

The masses for $X^A_i$, $A=10,45$, arose from $\lambda_i^A \;X_i^A X_i^A \phi$ and for simplicity we took $\langle \phi \rangle = V$, the scale of intermediate gauge symmetry breaking. If instead we take the $X^A_i$ masses to have the more general form $\lambda_i^A \;X_i^A X_i^A M^A$ we find that there is a single new parameter that enters the flavor observables, $R = M^{10}/M^{45}$. Furthermore, to accommodate $t$ and $b, \tau$ masses requires $R<100$.

In Fig. \ref{fig:fitwithYB} we show results for a fit that minimizes $d^2 = \sum_i \ln^2(O_i/O_{i,obs})$ for 15 observables, including $Y_B$, in terms of the 6 parameters $(\epsilon_i, \epsilon_X,V, R)$, in the MAFS approximation.  The fit is remarkably successful: 13 of the observables are accounted for at the factor of 2 level or better, including $Y_B$, while $m_u$ and $m_d$ are outliers, each differing from their observed value by about a factor of 4.  Furthermore, this fit raises the central MAFS value of $\bar{\nu}_2$ above the $10^9$ GeV scale needed for successful leptogenesis. The values of $d^2$ for this fit are shown in Table \ref{tab:chi_models}. Also shown are values of $d^2$ for the fits without $Y_B$, shown in Figure \ref{fig:SU5fit} for $SU(5)$ and Figure \ref{fig:fitnoYB} for $SO(10)$. 

\begin{table}[h!]
\centering
\renewcommand{\arraystretch}{1.3}
\begin{tabular}{|c|c|c|c|c|c|}
\hline
Model  & $N_{\rm obs}$ & $N_{\rm par}$ & $N_{\rm dof}$ & $d^2$ & $d^2_{\rm dof}$ \\
\hline
$SU(5)$ & $14$ & $7$ & $7$ &
$3.6$ &
$0.51$
\\
\hline
$SO(10)$  without  $Y_B$& $14$ & $5$ & $9$ &
$4.5$ &
$0.5$
\\
\hline
$SO(10)$ with $Y_B$ & $15$ & $6$ & $9$ &
$6.3$ &
$0.7$
\\
\hline
\end{tabular}
\caption{
Fits in the MAFS approximation in $SU(5)$ and in $SO(10)$, with and without $Y_B$. $N_{\rm obs}$ is the number of observables in the fit, $N_{\rm par}$ is the number of free parameters, $N_{\rm dof}=N_{\rm obs}-N_{\rm par}$ is the number of degrees of freedom, and $d^2_{\rm dof}=d^2/N_{\rm dof}$ is the d-square per degree of freedom.
}
\label{tab:chi_models}
\end{table}

\section{Conclusions}
\label{sec:concl}

We have introduced a framework involving Maximal Abelian Flavor Symmetries to describe, at the level of factors of two, the overall structure of the quark and lepton masses and mixings. When the gauge group is that of the Standard Model, all of the observed quark and lepton masses and mixings are consistent with MAFS \footnote{MAFS provides an approximate numerical reference value for the SM Yukawa coupling matrices, $Y^{U,D}_{ij} \approx \epsilon^q_i \epsilon^{\bar{u}, \bar{d}}_j$ and $Y^E_{ij} \approx \epsilon^\ell_i \epsilon^{\bar{e}}_j$, with $\epsilon^{q, {\bar u}, \bar{d}, \ell, \bar{e}}_i$ from Table \ref{tab:321eps}. This allows a convenient definition of a texture zero: in a theory of flavor, if any element is much smaller than its MAFS reference value it is a texture zero.}; perhaps not surprising, as there are 15 symmetry-breaking parameters $\epsilon^{q, {\bar u}, \bar{d}, \ell, \bar{e}}_i$.  The neutrino masses are required to have normal ordering and the observed modest mass hierarchy and large neutrino mixing angles are all consistent with very little hierarchy amongst the $\epsilon^{\ell}_i$. There is a single prediction $V_{ub}, V_{td} \approx V_{us} V_{cb}$, which is well satisfied. MAFS becomes much more powerful as the gauge symmetry is increased so that the quarks and leptons fit into fewer multiplets.

In the $SU(5)$ grand unified theory, with three generations in $T(10) + \bar{F}(\bar{5})$ multiplets, quark and lepton masses and mixings are described by just 6 MAFS parameters, $\epsilon^{T, \bar{F}}_i$, and the mass scale of lepton number violation, $M$. Remarkably, the 18 CP-conserving observables are well described by these, implying 11 successful approximate predictions from MAFS, shown in (\ref{eq:11pred}). The predictions are all within a factor of 2 of the data, with the exception of $m_\mu$  and $V_{us}$, which are off by about a factor of 3, as illustrated by the fit of Fig. \ref{fig:SU5fit}. Color factors of 3 can improve
the fit to $y_e/y_d$ and $y_\mu/y_s$ \cite{Georgi:1979df}; but even without them, the MAFS approximation works remarkably well in SU(5), and is insensitive to the scalar sector of the theory.

This significant success follows from the MAFS scaling behavior
\begin{equation}
         \frac{m_{u_i}}{m_{u_j}} \; \approx \; \left( \frac{\epsilon^T_i}{\epsilon^T_j} \right)^2, \hspace{0.4in} 
         \frac{m_{d_i}}{m_{d_j}}, \frac{m_{e_i}}{m_{e_j}} \; \approx \; \left( \frac{\epsilon^T_i}{\epsilon^T_j} \right) \left(\frac{\epsilon^{\bar{F}}_i}{\epsilon^{\bar{F}}_j} \right), \hspace{0.4in} 
         \frac{m_{\nu_i}}{m_{\nu_j}} \; \approx \; \left( \frac{\epsilon^{\bar{F}}_i}{\epsilon^{\bar{F}}_j} \right)^2.
\label{eq:5masshier}
\end{equation}
which correctly predicts the mass hierarchies in the $D/E$ sectors to be the geometric mean of those in the $U$ and $\nu$ sectors.
Furthermore, since the mass hierarchies in the $U$ sector are observed to be approximately the square of those in the $D$ and $E$ sectors, all the hierarchies arise mainly from $\epsilon^T_i/\epsilon^T_j$; the ratios $\epsilon^{\bar{F}}_i / \epsilon^{\bar{F}}_j$ are not far from unity. Indeed, MAFS in SU(5) predicts a very small hierarchy in neutrino masses and leads to the surprising conclusion that, even though quarks and leptons are unified, the neutrino mixing angles are large while quark mixing angles are small
\begin{equation}
         V_{ij} \; \approx \;\frac{\epsilon^T_i}{\epsilon^T_j} \ll1, \qquad
         U_{ij} \; \approx \;\frac{\epsilon^{\bar{F}}_i}{\epsilon^{\bar{F}}_j} \simeq {\cal O}(1), \qquad i<j.
\label{eq:5masshier}
\end{equation}
The scale of neutrino masses fixes 
\begin{align}
    \label{eq:MinSU5}
    M \; \approx \; \frac{y^2_b}{y_t} \frac{v^2}{m_{\nu_3}} \; \simeq 10^{11} \; \rm{GeV} 
\end{align}

In SU(5), flavor can be broadly described by 5 small flavor parameters, $(\epsilon^T_{1,2}, \epsilon^{\bar{F}}_i)$. In a simple $SU(5)$ theory in 5 dimensions, with all fermions propagating in the bulk except $T_3$, all flavor hierarchies of the charged fermions of the heaviest two generations can be broadly understood from a single volume factor, $\epsilon_V \sim 0.03$. Including the lightest generation, quark and charged lepton masses can be understood in terms of just 3 small parameters, $\epsilon_V \sim 0.03, \,\epsilon^{T, \bar{F}}_1 \sim 0.07$. 

In $SO(10)$ unified theories, with minimal fermion multiplets $\psi_i(\bf{16})$, the observed flavor hierarchies are inconsistent with MAFS: for example, the ratio $y_t/y_b$ cannot be understood, and there is no understanding of why mixing angles are much larger for leptons than quarks. This can be remedied by introducing a single additional fermion multiplet, $X$, with $SO(10)$ breaking inducing $\psi_3 / X$ mass mixing so that $(\bar{d}_3, \ell_3)$ lie mainly in $X$ while $(q_3, \bar{u}_3, \bar{e}_3, \bar{\nu}_3)$ lie mainly in $\psi_3$. In this case, MAFS can account approximately for all flavor observables in terms of the 4 parameters $\epsilon_1 \sim 0.002, \epsilon_2 \sim 0.02, \epsilon_3 \sim 0.6, \epsilon_X \sim 0.01$, as shown by the fit in Figure \ref{fig:fitnoYB}. Thus, the remarkable understanding of a generation in terms of the spinor $\bf{16}$ of $SO(10)$ is found to be compatible with a very economic description of the 17 observed flavor parameters. 

This flavor structure is independent of the scalar sector of the theory, except that $SO(10)$ must first break to some intermediate gauge group which breaks to the SM gauge group via a vev $V \approx 10^{12}$ GeV to account for the overall mass scale of the neutrinos. To be specific, we focused on an $SO(10)$ model with the SM Higgs embedded in spinor representations. This leads to a unified seesaw description of all masses in the $U,D,E$ and $\nu$ sectors, except for $(b, \tau)$ which arise from $\psi_3/X$ mixing. A minimal version of this theory has been studied in detail \cite{Carrasco-Martinez:2025zus}; it is consistent with precision gauge coupling unification and proton decay, predicts the SM Higgs mass and solves the strong CP problem using spacetime parity. 

In any theory with very heavy right-handed neutrinos, the cosmological baryon asymmetry may result from leptogenesis. Can MAFS give an approximate prediction for the baryon asymmetry? With the SM gauge group, the observed baryon asymmetry is consistent with MAFS but cannot be predicted, because it depends on $\epsilon$ parameters that are not determined.  In the $SU(5)$ theory, a correct prediction for the order of magnitude of the cosmological baryon asymmetry results if the MAFS parameters on the relevant right-handed neutrinos and the relevant CP violating phase are order unity (\ref{eq:YBmax}). Furthermore, the prediction is almost independent of $\epsilon^{T, \bar{F}}_i$
\begin{align}
    \label{eq:YBmax2}
    Y_B(SU(5)) \;\approx \; \frac{42}{79} Y_{\rm therm} \;  \frac{M}{M_{Pl}}  \; \approx \; \frac{42}{79} Y_{\rm therm}  \; \frac{m^2_b \,v}{m_t \,m_{\nu_3} \,M_{Pl}} \; \approx \;   10^{-10}. 
\end{align}

The case of $SO(10)$ is extremely interesting because there are no additional $\epsilon_a$ parameters associated with $\bar{\nu}_i$, since these lie in $\psi_i(\bf{16})$. The baryon asymmetry can be estimated in the MAFS approximation from measured flavor observables. We find this estimate is too low by two orders of magnitude.  This discrepancy may result from a pileup of the unknown order unity coefficients that MAFS ignores, or it may be that an assumption must be relaxed.  We have assumed that the mass scale associated with the $X$ state arises from the scale of intermediate gauge symmetry breaking; if we relax this by a factor of $4$ we find an excellent fit to both flavor observables and $Y_B$, as shown in Figure \ref{fig:fitwithYB}. 

\section*{Acknowledgements}
LJH thanks Riccardo Barbieri, Simon Knapen, Zoltan Ligeti and Claudio Manzari for conversations and comments. This work was supported by the NSF grant PHY-2515115 and the Office of High Energy Physics of the U.S. Department of
Energy under contract DE-AC02-05CH11231.


\appendix

\section{MAFS Test for Texture Zeros}
\label{sec:texture0}

Many theories of flavor lead to ``texture zeros'', matrix elements that are zero or negligibly small. This results in Froggatt-Nielsen theories and with non-Abelian flavor symmetries, such as $U(2)$, $S_3$ \cite{Hall:1995es}, and $A_4$.  Even if an entry vanishes at leading order, it will typically get generated at higher order, leading to important questions. What does ``negligibly small'' mean? At what level do such small terms affect predictions for observables? MAFS gives remarkably simple answers to these questions. While the flavor group in MAFS is large, it does not lead to precise predictions because it is completely broken. To avoid massless fermions all $\epsilon_a$ must be non-zero, so that all gauge-invariant flavor interactions of (\ref{eq:MAFSinteraction}) are non-zero; there are no texture zeros

In any theory, consider a particular Yukawa interaction $Y_{ab} \; \psi_a \psi_b \phi$, where $\psi_{a,b}$ contain SM fermions.  When MAFS is applied to this theory, values for $\epsilon_{a,b}$ are derived by fitting to flavor observables. If there are more $\epsilon_a$ than observables, $\epsilon_a$ are chosen to be as large as possible such that no tuned cancellations between terms are needed to describe the observables. The MAFS estimation for this coupling is $|Y_{ab}| \approx \epsilon_a \epsilon_b$. With this size, the coupling has order unity effects on at least some flavor observables. It may be that the theory is consistent with data even if this particular coupling vanishes. If $|Y_{ab}| \ll \epsilon_a \epsilon_b$ this entry becomes a texture zero, and the fractional effect it has on relevant observables $O$ is 
\begin{equation}
\frac{\delta O}{O} \sim \; \frac{|Y_{ab}|}{\epsilon_a \epsilon_b}.
\label{eq:texture0example}
\end{equation}
While this can be applied to more than one texture zero of a theory, there is a limit to how many texture zeros allow a realistic fit to the data.

As an example, consider applying MAFS to the SM.  The Yukawa matrices are estimated using $Y^{U,D}_{ij} \approx \epsilon^q_i \epsilon^{\bar{u}, \bar{d}}_j$ and $Y^E_{ij} \approx \epsilon^\ell_i \epsilon^{\bar{e}}_j$, using $\epsilon^{q, {\bar u}, \bar{d}, \ell, \bar{e}}_i$ from Table \ref{tab:321eps}.  This gives a numerical reference for all entries of the SM Yukawa coupling matrices.  If some BSM theory of flavor gives rise to some matrix element much less than the reference value then that entry is a texture zero.  Furthermore, the suppression relative to the MAFS reference value gives an estimate of the perturbation from this small entry on the flavor observables. For example,
\begin{equation}
\frac{|Y^U_{12}|}{ \epsilon^q_1 \epsilon^{\bar{u}}_2} \; = \; \delta \ll 1 
\qquad \implies \qquad \frac{\delta y_u}{y_u}, \; \frac{\delta V_{us}}{V_{us}} \; \sim \; \delta.
\label{eq:texture0example}
\end{equation}

\bibliographystyle{JHEP}
\bibliography{biblio}

@article{Greljo:2025mwj,
    author = "Greljo, Admir and Palavri{\'c}, Ajdin and Stefanek, Ben A.",
    title = "{Minimal Flavor Protection for TeV-scale New Physics}",
    eprint = "2512.04159",
    archivePrefix = "arXiv",
    primaryClass = "hep-ph",
    month = "12",
    year = "2025"
}

@article{Carone:1997qg,
    author = "Carone, Christopher D. and Hall, Lawrence J.",
    title = "{Neutrino physics from a U(2) flavor symmetry}",
    eprint = "hep-ph/9702430",
    archivePrefix = "arXiv",
    reportNumber = "LBL-40024, LBNL-40024, UCB-PTH-97-08",
    doi = "10.1103/PhysRevD.56.4198",
    journal = "Phys. Rev. D",
    volume = "56",
    pages = "4198--4206",
    year = "1997"
}

@article{Barbieri:1999pe,
    author = "Barbieri, Riccardo and Creminelli, Paolo and Romanino, Andrea",
    title = "{Neutrino mixings from a U(2) flavor symmetry}",
    eprint = "hep-ph/9903460",
    archivePrefix = "arXiv",
    reportNumber = "SNS-PH-99-1, OUTP-99-19-P",
    doi = "10.1016/S0550-3213(99)00425-3",
    journal = "Nucl. Phys. B",
    volume = "559",
    pages = "17--26",
    year = "1999"
}

@article{Antusch:2023shi,
    author = "Antusch, Stefan and Greljo, Admir and Stefanek, Ben A. and Thomsen, Anders Eller",
    title = "{U(2) Is Right for Leptons and Left for Quarks}",
    eprint = "2311.09288",
    archivePrefix = "arXiv",
    primaryClass = "hep-ph",
    reportNumber = "KCL-PH-TH/2023-64",
    doi = "10.1103/PhysRevLett.132.151802",
    journal = "Phys. Rev. Lett.",
    volume = "132",
    number = "15",
    pages = "151802",
    year = "2024"
}

@article{Ibe:2024cvi,
    author = "Ibe, Masahiro and Shirai, Satoshi and Watanabe, Keiichi",
    title = "{Comprehensive Bayesian exploration of Froggatt-Nielsen mechanism}",
    eprint = "2412.19484",
    archivePrefix = "arXiv",
    primaryClass = "hep-ph",
    reportNumber = "IPMU24-0047",
    doi = "10.1007/JHEP03(2025)150",
    journal = "JHEP",
    volume = "03",
    pages = "150",
    year = "2025"
}

@article{Cornella:2023zme,
    author = "Cornella, Claudia and Curtin, David and Neil, Ethan T. and Thompson, Jedidiah O.",
    title = "{Mapping and probing Froggatt-Nielsen solutions to the quark flavor puzzle}",
    eprint = "2306.08026",
    archivePrefix = "arXiv",
    primaryClass = "hep-ph",
    reportNumber = "MITP-23-026",
    doi = "10.1103/PhysRevD.111.015042",
    journal = "Phys. Rev. D",
    volume = "111",
    number = "1",
    pages = "015042",
    year = "2025"
}

@article{Altarelli:2005yx,
    author = "Altarelli, Guido and Feruglio, Ferruccio",
    title = "{Tri-bimaximal neutrino mixing, A(4) and the modular symmetry}",
    eprint = "hep-ph/0512103",
    archivePrefix = "arXiv",
    reportNumber = "CERN-PH-TH-2005-226",
    doi = "10.1016/j.nuclphysb.2006.02.015",
    journal = "Nucl. Phys. B",
    volume = "741",
    pages = "215--235",
    year = "2006"
}

@article{Ma:2001dn,
    author = "Ma, Ernest and Rajasekaran, G.",
    title = "{Softly broken A(4) symmetry for nearly degenerate neutrino masses}",
    eprint = "hep-ph/0106291",
    archivePrefix = "arXiv",
    reportNumber = "UCRHEP-T308",
    doi = "10.1103/PhysRevD.64.113012",
    journal = "Phys. Rev. D",
    volume = "64",
    pages = "113012",
    year = "2001"
}

@article{Pomarol:1995xc,
    author = "Pomarol, Alex and Tommasini, Daniele",
    title = "{Horizontal symmetries for the supersymmetric flavor problem}",
    eprint = "hep-ph/9507462",
    archivePrefix = "arXiv",
    reportNumber = "CERN-TH-95-207",
    doi = "10.1016/0550-3213(96)00074-0",
    journal = "Nucl. Phys. B",
    volume = "466",
    pages = "3--24",
    year = "1996"
}

@article{King:2001uz,
    author = "King, S. F. and Ross, Graham G.",
    title = "{Fermion masses and mixing angles from SU(3) family symmetry}",
    eprint = "hep-ph/0108112",
    archivePrefix = "arXiv",
    reportNumber = "SHEP-01-21, OUTP-01-46P",
    doi = "10.1016/S0370-2693(01)01139-X",
    journal = "Phys. Lett. B",
    volume = "520",
    pages = "243--253",
    year = "2001"
}

@article{Barbieri:1995uv,
    author = "Barbieri, Riccardo and Dvali, G. R. and Hall, Lawrence J.",
    title = "{Predictions from a U(2) flavor symmetry in supersymmetric theories}",
    eprint = "hep-ph/9512388",
    archivePrefix = "arXiv",
    reportNumber = "LBL-38065, UCB-PTH-95-44",
    doi = "10.1016/0370-2693(96)00318-8",
    journal = "Phys. Lett. B",
    volume = "377",
    pages = "76--82",
    year = "1996"
}

@article{Faroughy:2020ina,
    author = "Faroughy, Darius A. and Isidori, Gino and Wilsch, Felix and Yamamoto, Kei",
    title = "{Flavour symmetries in the SMEFT}",
    eprint = "2005.05366",
    archivePrefix = "arXiv",
    primaryClass = "hep-ph",
    doi = "10.1007/JHEP08(2020)166",
    journal = "JHEP",
    volume = "08",
    pages = "166",
    year = "2020"
}

@article{Arkani-Hamed:1999ylh,
    author = "Arkani-Hamed, Nima and Schmaltz, Martin",
    title = "{Hierarchies without symmetries from extra dimensions}",
    eprint = "hep-ph/9903417",
    archivePrefix = "arXiv",
    reportNumber = "SLAC-PUB-8082",
    doi = "10.1103/PhysRevD.61.033005",
    journal = "Phys. Rev. D",
    volume = "61",
    pages = "033005",
    year = "2000"
}

@article{Froggatt:1978nt,
    author = "Froggatt, C. D. and Nielsen, Holger Bech",
    title = "{Hierarchy of Quark Masses, Cabibbo Angles and CP Violation}",
    reportNumber = "CERN-TH-2519",
    doi = "10.1016/0550-3213(79)90316-X",
    journal = "Nucl. Phys. B",
    volume = "147",
    pages = "277--298",
    year = "1979"
}

@article{Leurer:1992wg,
    author = "Leurer, Miriam and Nir, Yosef and Seiberg, Nathan",
    title = "{Mass matrix models}",
    eprint = "hep-ph/9212278",
    archivePrefix = "arXiv",
    reportNumber = "RU-92-59, WIS-92-94-PH",
    doi = "10.1016/0550-3213(93)90112-3",
    journal = "Nucl. Phys. B",
    volume = "398",
    pages = "319--342",
    year = "1993"
}

@article{Leurer:1993gy,
    author = "Leurer, Miriam and Nir, Yosef and Seiberg, Nathan",
    title = "{Mass matrix models: The Sequel}",
    eprint = "hep-ph/9310320",
    archivePrefix = "arXiv",
    reportNumber = "RU-93-43, WIS-93-93-PH",
    doi = "10.1016/0550-3213(94)90074-4",
    journal = "Nucl. Phys. B",
    volume = "420",
    pages = "468--504",
    year = "1994"
}

@article{Hall:1995es,
    author = "Hall, Lawrence J. and Murayama, Hitoshi",
    title = "{A Geometry of the generations}",
    eprint = "hep-ph/9508296",
    archivePrefix = "arXiv",
    reportNumber = "UCB-PTH-95-29, LBL-37627",
    doi = "10.1103/PhysRevLett.75.3985",
    journal = "Phys. Rev. Lett.",
    volume = "75",
    pages = "3985--3988",
    year = "1995"
}

@article{Hall:1993ca,
    author = "Hall, Lawrence J. and Weinberg, Steven",
    title = "{Flavor changing scalar interactions}",
    eprint = "hep-ph/9303241",
    archivePrefix = "arXiv",
    reportNumber = "UTTG-22-92, LBL-33016, UCB-PTH-92-36",
    doi = "10.1103/PhysRevD.48.R979",
    journal = "Phys. Rev. D",
    volume = "48",
    pages = "R979--R983",
    year = "1993"
}

@article{Antaramian:1992ya,
    author = "Antaramian, Aram and Hall, Lawrence J. and Rasin, Andrija",
    title = "{Flavor changing interactions mediated by scalars at the weak scale}",
    eprint = "hep-ph/9206205",
    archivePrefix = "arXiv",
    reportNumber = "LBL-32345, UCB-PTH-92-18",
    doi = "10.1103/PhysRevLett.69.1871",
    journal = "Phys. Rev. Lett.",
    volume = "69",
    pages = "1871--1873",
    year = "1992"
}

@article{Carrasco-Martinez:2025zus,
    author = "Carrasco-Martinez, Juanca and Hall, Lawrence J. and Harigaya, Keisuke and Langhoff, Kevin",
    title = "{A flavor of SO(10) unification with a spinor Higgs}",
    eprint = "2506.20708",
    archivePrefix = "arXiv",
    primaryClass = "hep-ph",
    doi = "10.1007/JHEP11(2025)073",
    journal = "JHEP",
    volume = "11",
    pages = "073",
    year = "2025"
}

@article{Kawamura:2000ev,
    author = "Kawamura, Yoshiharu",
    title = "{Triplet doublet splitting, proton stability and extra dimension}",
    eprint = "hep-ph/0012125",
    archivePrefix = "arXiv",
    reportNumber = "DPSU-00-03",
    doi = "10.1143/PTP.105.999",
    journal = "Prog. Theor. Phys.",
    volume = "105",
    pages = "999--1006",
    year = "2001"
}

@article{Hall:2002ci,
    author = "Hall, Lawrence J. and Nomura, Yasunori",
    title = "{A Complete theory of grand unification in five-dimensions}",
    eprint = "hep-ph/0205067",
    archivePrefix = "arXiv",
    reportNumber = "UCB-PTH-02-20, LBNL-50199",
    doi = "10.1103/PhysRevD.66.075004",
    journal = "Phys. Rev. D",
    volume = "66",
    pages = "075004",
    year = "2002"
}

@article{Hebecker:2001wq,
    author = "Hebecker, Arthur and March-Russell, John",
    title = "{A Minimal S**1 / (Z(2) x Z-prime (2)) orbifold GUT}",
    eprint = "hep-ph/0106166",
    archivePrefix = "arXiv",
    reportNumber = "CERN-TH-2001-156, LBNL-48220",
    doi = "10.1016/S0550-3213(01)00374-1",
    journal = "Nucl. Phys. B",
    volume = "613",
    pages = "3--16",
    year = "2001"
}

@article{Altarelli:2001qj,
    author = "Altarelli, Guido and Feruglio, Ferruccio",
    title = "{SU(5) grand unification in extra dimensions and proton decay}",
    eprint = "hep-ph/0102301",
    archivePrefix = "arXiv",
    reportNumber = "CERN-TH-2001-028, DFPD-01-TH-04",
    doi = "10.1016/S0370-2693(01)00650-5",
    journal = "Phys. Lett. B",
    volume = "511",
    pages = "257--264",
    year = "2001"
}

@article{Hall:2001pg,
    author = "Hall, Lawrence J. and Nomura, Yasunori",
    title = "{Gauge unification in higher dimensions}",
    eprint = "hep-ph/0103125",
    archivePrefix = "arXiv",
    reportNumber = "UCB-PTH-01-08, LBNL-47610",
    doi = "10.1103/PhysRevD.64.055003",
    journal = "Phys. Rev. D",
    volume = "64",
    pages = "055003",
    year = "2001"
}

@article{Yanagida:1998jk,
    author = "Yanagida, T. and Sato, J.",
    editor = "Suzuki, Y. and Totsuka, Y.",
    title = "{Large lepton mixing in seesaw models: Coset space family unification}",
    eprint = "hep-ph/9809307",
    archivePrefix = "arXiv",
    doi = "10.1016/S0920-5632(99)00431-4",
    journal = "Nucl. Phys. B Proc. Suppl.",
    volume = "77",
    pages = "293--298",
    year = "1999"
}

@article{Buchmuller:1998zf,
    author = "Buchmuller, W. and Yanagida, T.",
    title = "{Quark lepton mass hierarchies and the baryon asymmetry}",
    eprint = "hep-ph/9810308",
    archivePrefix = "arXiv",
    reportNumber = "DESY-98-155",
    doi = "10.1016/S0370-2693(98)01480-4",
    journal = "Phys. Lett. B",
    volume = "445",
    pages = "399--402",
    year = "1999"
}

@article{Altarelli:1998ns,
    author = "Altarelli, Guido and Feruglio, Ferruccio",
    title = "{A Simple grand unification view of neutrino mixing and fermion mass matrices}",
    eprint = "hep-ph/9812475",
    archivePrefix = "arXiv",
    reportNumber = "CERN-TH-98-410, DFPD-98-TH-52",
    doi = "10.1016/S0370-2693(99)00208-7",
    journal = "Phys. Lett. B",
    volume = "451",
    pages = "388--396",
    year = "1999"
}

@article{Babu_1996,
   title={Large neutrino mixing angles in unified theories},
   volume={381},
   ISSN={0370-2693},
   url={http://dx.doi.org/10.1016/0370-2693(96)00552-7},
   DOI={10.1016/0370-2693(96)00552-7},
   number={1–3},
   journal={Physics Letters B},
   publisher={Elsevier BV},
   author={Babu, K.S and Barr, S.M},
   year={1996},
   month=jul, pages={202–208} }

@article{Babu:1995uu,
    author = "Babu, K. S. and Barr, Stephen M.",
    title = "{Realistic quark and lepton masses through SO(10) symmetry}",
    eprint = "hep-ph/9512389",
    archivePrefix = "arXiv",
    reportNumber = "IASSNS-HEP-95-99, BA-95-55",
    doi = "10.1103/PhysRevD.56.2614",
    journal = "Phys. Rev. D",
    volume = "56",
    pages = "2614--2631",
    year = "1997"
}

@article{Georgi:1974my,
    author = "Georgi, Howard",
    editor = "Carlson, Hugh C. Carl E. Wolfe",
    title = "{The State of the Art\textemdash{}Gauge Theories}",
    doi = "10.1063/1.2947450",
    journal = "AIP Conf. Proc.",
    volume = "23",
    pages = "575--582",
    year = "1975"
}

@article{Fukugita:1986hr,
	Author = {Fukugita, M. and Yanagida, T.},
	Doi = {10.1016/0370-2693(86)91126-3},
	Journal = {Phys. Lett.},
	Pages = {45-47},
	Reportnumber = {RIFP-641},
	Slaccitation = {%%CITATION = PHLTA,B174,45;%%},
	Title = {{Baryogenesis Without Grand Unification}},
	Volume = {B174},
	Year = {1986}}

@article{Hall:2018let,
	Archiveprefix = {arXiv},
	Author = {Hall, Lawrence J. and Harigaya, Keisuke},
	Doi = {10.1007/JHEP10(2018)130},
	Eprint = {1803.08119},
	Journal = {JHEP},
	Pages = {130},
	Primaryclass = {hep-ph},
	Slaccitation = {%%CITATION = ARXIV:1803.08119;%%},
	Title = {{Implications of Higgs Discovery for the Strong CP Problem and Unification}},
	Volume = {10},
	Year = {2018}}

@article{Hall:2019qwx,
	Archiveprefix = {arXiv},
	Author = {Hall, Lawrence J. and Harigaya, Keisuke},
	Doi = {10.1007/JHEP11(2019)033},
	Eprint = {1905.12722},
	Journal = {JHEP},
	Pages = {033},
	Primaryclass = {hep-ph},
	Slaccitation = {%%CITATION = ARXIV:1905.12722;%%},
	Title = {{Higgs Parity Grand Unification}},
	Volume = {11},
	Year = {2019}}

@article{Fritzsch:1974nn,
    author = "Fritzsch, Harald and Minkowski, Peter",
    doi = "10.1016/0003-4916(75)90211-0",
    journal = "Annals Phys.",
    pages = "193--266",
    reportNumber = "CALT-68-467",
    title = "{Unified Interactions of Leptons and Hadrons}",
    volume = "93",
    year = "1975"
}

@article{Georgi:1974sy,
    author = "Georgi, H. and Glashow, S. L.",
    title = "{Unity of All Elementary Particle Forces}",
    doi = "10.1103/PhysRevLett.32.438",
    journal = "Phys. Rev. Lett.",
    volume = "32",
    pages = "438--441",
    year = "1974"
}

@article{Georgi:1979df,
    author = "Georgi, Howard and Jarlskog, C.",
    title = "{A New Lepton - Quark Mass Relation in a Unified Theory}",
    reportNumber = "HUTP-79-A026",
    doi = "10.1016/0370-2693(79)90842-6",
    journal = "Phys. Lett. B",
    volume = "86",
    pages = "297--300",
    year = "1979"
}

@article{Giudice:2003jh,
    author = "Giudice, G. F. and Notari, A. and Raidal, M. and Riotto, A. and Strumia, A.",
    title = "{Towards a complete theory of thermal leptogenesis in the SM and MSSM}",
    eprint = "hep-ph/0310123",
    archivePrefix = "arXiv",
    reportNumber = "IFUP-TH-2003-37, CERN-TH-2003-240",
    doi = "10.1016/j.nuclphysb.2004.02.019",
    journal = "Nucl. Phys. B",
    volume = "685",
    pages = "89--149",
    year = "2004"
}

@article{Dolinski:2019nrj,
    author = "Dolinski, Michelle J. and Poon, Alan W. P. and Rodejohann, Werner",
    title = "{Neutrinoless Double-Beta Decay: Status and Prospects}",
    eprint = "1902.04097",
    archivePrefix = "arXiv",
    primaryClass = "nucl-ex",
    doi = "10.1146/annurev-nucl-101918-023407",
    journal = "Ann. Rev. Nucl. Part. Sci.",
    volume = "69",
    pages = "219--251",
    year = "2019"
}

@article{nEXO:2021ujk,
    author = "Adhikari, G. and others",
    collaboration = "nEXO",
    title = "{nEXO: neutrinoless double beta decay search beyond 10$^{28}$ year half-life sensitivity}",
    eprint = "2106.16243",
    archivePrefix = "arXiv",
    primaryClass = "nucl-ex",
    doi = "10.1088/1361-6471/ac3631",
    journal = "J. Phys. G",
    volume = "49",
    number = "1",
    pages = "015104",
    year = "2022"
}

@article{LEGEND:2021bnm,
    author = "Abgrall, N. and others",
    collaboration = "LEGEND",
    title = "{The Large Enriched Germanium Experiment for Neutrinoless $\beta\beta$ Decay}: {LEGEND-1000 Preconceptual Design Report}",
    eprint = "2107.11462",
    archivePrefix = "arXiv",
    primaryClass = "physics.ins-det",
    month = "7",
    year = "2021"
}

@article{Huang:2020hdv,
    author = "Huang, Guo-yuan and Zhou, Shun",
    title = "{Precise Values of Running Quark and Lepton Masses in the Standard Model}",
    eprint = "2009.04851",
    archivePrefix = "arXiv",
    primaryClass = "hep-ph",
    doi = "10.1103/PhysRevD.103.016010",
    journal = "Phys. Rev. D",
    volume = "103",
    number = "1",
    pages = "016010",
    year = "2021"
}

@article{Elwood:1998kf,
    author = "Elwood, John K. and Irges, Nikolaos and Ramond, Pierre",
    title = "{Family symmetry and neutrino mixing}",
    eprint = "hep-ph/9807228",
    archivePrefix = "arXiv",
    reportNumber = "UFIFT-HEP-98-14",
    doi = "10.1103/PhysRevLett.81.5064",
    journal = "Phys. Rev. Lett.",
    volume = "81",
    pages = "5064--5067",
    year = "1998"
}

@article{1Dimopoulos:1983,
  author = {S. Dimopoulos},
  title = {Natural generation of fermion masses},
  doi = "10.1016/0370-2693(83)90132-6",
  journal = {Phys. Lett. B},
  volume = {129},
  pages = {417--428},
  year = {1983}
}

@article{2Berezhiani:1983,
   author = "Berezhiani, Z. G.",
    title = "{The Weak Mixing Angles in Gauge Models with Horizontal Symmetry: A New Approach to Quark and Lepton Masses}",
    doi = "10.1016/0370-2693(83)90737-2",
    journal = "Phys. Lett. B",
    volume = "129",
    pages = "99--102",
    year = "1983"
}

@inproceedings{3Bagger:1984,
   author = "Bagger, Jonathan and Dimopoulos, Savas and Georgi, Howard and Raby, Stuart",
    title = "{THEORIES OF FERMION MASSES}",
    booktitle = "{Fifth Workshop on Grand Unification}",
    reportNumber = "SLAC-PUB-3342",
    month = "5",
    year = "1984"
}

@article{4Davidson:1984,
    author = "Davidson, Aharon and Nair, V. P. and Wali, Kameshwar C.",
    title = "{Mixing Angles and {CP} Violation in the SO(10) X U(1)-(pq) Model}",
    reportNumber = "SU-4222-265, COO-3533-265",
    doi = "10.1103/PhysRevD.29.1513",
    journal = "Phys. Rev. D",
    volume = "29",
    pages = "1513",
    year = "1984"
}

@article{5Bagger:1985,
   author = "Bagger, Jonathan A. and Dimopoulos, Savas and Masso, Eduard and Reno, M. Hall",
    title = "{A realistic theory of family unification}",
    reportNumber = "SLAC-PUB-3441",
    doi = "10.1016/0550-3213(85)90627-3",
    journal = "Nucl. Phys. B",
    volume = "258",
    pages = "565--600",
    year = "1985"
}

@article{6Berezhiani:1985,
    author = "Berezhiani, Z. G.",
    title = "{Horizontal Symmetry and Quark - Lepton Mass Spectrum: The SU(5) x SU(3)-h Model}",
    doi = "10.1016/0370-2693(85)90164-9",
    journal = "Phys. Lett. B",
    volume = "150",
    pages = "177--181",
    year = "1985"
}

@article{7Davidson:1988,
  author = {A. Davidson and K. C. Wali},
    author = "Davidson, Aharon and Wali, Kameshwar C.",
    title = "{Family Mass Hierarchy From Universal Seesaw Mechanism}",
    reportNumber = "SU-4228-375",
    doi = "10.1103/PhysRevLett.60.1813",
    journal = "Phys. Rev. Lett.",
    volume = "60",
    pages = "1813",
    year = "1988"
}

@article{8Davidson:1990,
    author = "Davidson, Aharon and Ranfone, Stefano and Wali, Kameshwar C.",
    title = "{Quark Masses and Mixing Angles From Universal Seesaw Mechanism}",
    reportNumber = "SU-4228-410",
    doi = "10.1103/PhysRevD.41.208",
    journal = "Phys. Rev. D",
    volume = "41",
    pages = "208",
    year = "1990"
}

@article{10Dimopoulos:1992b,
   author = "Dimopoulos, Savas and Hall, Lawrence J. and Raby, Stuart",
    title = "{A Predictive ansatz for fermion mass matrices in supersymmetric grand unified theories}",
    reportNumber = "UCB-PTH-91-61, LBL-31441, DOE-ER-10545-567, ITP-908-91-STANFORD",
    doi = "10.1103/PhysRevD.45.4192",
    journal = "Phys. Rev. D",
    volume = "45",
    pages = "4192--4200",
    year = "1992"
}

@article{9Dimopoulos:1992,
    author = "Dimopoulos, Savas and Hall, Lawrence J. and Raby, Stuart",
    title = "{A Predictive framework for fermion masses in supersymmetric theories}",
    reportNumber = "UCB-PTH-91-60, LBL-31440, DOE-ER-01545-566, ITP-907-91-STANFORD",
    doi = "10.1103/PhysRevLett.68.1984",
    journal = "Phys. Rev. Lett.",
    volume = "68",
    pages = "1984--1987",
    year = "1992"
}

\end{document}